\documentclass[12pt]{iopart}
\usepackage{graphicx}
\usepackage{bm}
\begin{document}
\title{Persistence and survival in 
equilibrium step fluctuations}
\author{M. Constantin$^1$, C. Dasgupta$^2$, S. Das Sarma$^3$, D. B.
Dougherty$^4$, and E. D. Williams$^5$}
\address{$^1$ Department of Physics and Astronomy, University of California, 
Irvine, CA 92697-4575, USA}
\ead{magda@uci.edu}
\address{$^2$ Centre for Condensed Matter Theory, 
Department of Physics, Indian Institute of Science, Bangalore 560012,
India}
\ead{cdgupta@physics.iisc.ernet.in}
\address{$^3$ Condensed Matter Theory Center, Department of Physics, University
of Maryland, College Park, MD 20742-4111, USA}
\ead{dassarma@physics.umd.edu}
\address{$^4$ Surface and Microanalysis Science Division, NIST, 100 Bureau
Dr., Mailstop 8372, Gaithersburg, MD 20899-8372, USA}
\ead{dougherty.daniel@gmail.com} 
\address{$^5$ Department of Physics and MRSEC, 
University of Maryland, College Park, MD 20742-4111, USA}
\ead{edw@umd.edu}
\begin{abstract}
Results of analytic and numerical investigations of first-passage
properties of equilibrium fluctuations of monatomic steps on a 
vicinal surface are reviewed. Both
temporal and spatial persistence and survival probabilities, as well as the 
probability of persistent large deviations are considered. Results of
experiments in which dynamical
scanning tunneling microscopy is used to evaluate these first-passage
properties for steps with different microscopic mechanisms of 
mass transport are also presented and interpreted in terms of
theoretical predictions for appropriate models. Effects of discrete
sampling, finite system size and finite observation time, which are 
important in understanding the results of experiments and simulations, are
discussed. 
\end{abstract}
\maketitle

\section{Introduction}
\label{intro}

Many problems in physics require understanding the 
stochastic dynamics of spatially extended objects. Traditionally, the
equilibrium and nonequilibrium dynamical properties of such systems 
have been described in terms of space- and time-dependent correlation
functions. In recent years, there has been considerable interest in 
studies of first-passage properties~\cite{redner_book,pers_review} of 
the dynamical fluctuations of such objects, quantified in terms of 
{\it persistence} and {\it survival} probabilities. The first-passage problem
for a temporally fluctuating quantity involves determining the distribution
of times for the quantity to cross a specified reference value {\it for the
first time}.
The persistence probability, defined as the probability of the 
stochastic variable {\it not} returning to its initial value in a specified
time interval, is essentially an integral of the distribution of the
corresponding first-passage time. A closely related quantity, the survival
probability, measures the probability of not crossing some other reference
point, such as the average value of the stochastic process, in a specified
time interval. These probabilities
depend on the whole history of the time evolution of the
system in the specified time interval, and 
provide a characterization of the 
underlying stochastic dynamics that is complementary to, 
and in some sense more detailed than a description 
based on correlation functions. 

Persistence and survival probabilities have been used in recent years
to describe the statistics of first-passage events in a variety of
spatially extended stochastic systems. Examples of such applications
range from the classical diffusion equation~\cite{diffusion} 
to the zero-temperature dynamics 
of ferromagnetic Ising and Potts models~\cite{spin1,spin2}, nonequilibrium
critical dynamics~\cite{critical}, 
reaction-diffusion processes in disordered environments~\cite{fisher},
and volatility in stock market fluctuations~\cite{stock}.  
An increasing
number of experimental results are also available for persistence 
and survival in systems such as
coalescence of droplets~\cite{droplet}, coarsening of two-dimensional soap 
froth~\cite{soap}, twisted
nematic liquid crystal~\cite{liqcryst}, nuclear spin distribution in 
laser-polarized atomic
gas~\cite{Xe}, and slow combustion front in paper~\cite{px_expt}.

There also has been much recent interest in experimental and theoretical
studies of the stochastic dynamics of growth and fluctuation of structures
on surfaces~\cite{book1,krugrev,book3,misbah}. The technique of imaging
spatial distributions and temporal variations of structures on surfaces
using scanning tunneling microscopy (STM) has made it possible to perform
experiments in which theoretical predictions can be
tested directly. Equilibrium fluctuations of crystal layer boundaries
or {\it steps} on a vicinal surface, obtained by cutting a crystal in a 
direction close to a high-symmetry plane, have been extensively 
studied~\cite{ellen,giesen} 
in this context. A vicinal surface consists of an array of monatomic
steps separated by terraces of the high-symmetry plane.
In thermal equilibrium, which can be achieved in experiments if the 
temperature is sufficiently high, 
the one-dimensional (1d) steps roughen due to thermal
fluctuations. 
The stochastic dynamics of these fluctuations is theoretically 
modeled by Langevin equations~\cite{krugrev,ellen,giesen} and atomistic,
solid-on-solid models~\cite{krugrev}. Theoretical and experimental 
investigations~\cite{ellen,giesen} of various space- and time-dependent
correlation functions of equilibrium step fluctuations have provided a 
wealth of information about the physical parameters that govern these
fluctuations.

In this paper, we review the results of our recent analytic, numerical and
experimental investigations
~\cite{Dan,alex,PTS,ST,STS,PT,PX,map,Dan2,expt3,ssp,spexp,unpub} of various
first-passage properties of equilibrium step fluctuations. In these
studies, both temporal
and spatial persistence and survival probabilities of different kinds are
considered and the results of STM experiments on systems with different 
microscopic mechanisms of step-edge fluctuations are compared directly with
the predictions of analytic and numerical calculations for appropriate
models. The first studies of first-passage statistics of surface growth
and fluctuations were carried out by Krug and 
co-workers~\cite{Krug1,Krug2} 
who obtained analytic and numerical results for the persistence probability
for several Langevin equations for interface dynamics. Our work
addresses many other interesting questions about the first-passage statistics
of interface fluctuations. The motivation for these studies
arises partly from the possibility of making direct comparisons between
experimental observations and theoretical predictions for various
first-passage properties of these systems.
Such comparisons provide an opportunity to validate the theoretical models
commonly used to describe step fluctuations. 
As noted in Ref.~\cite{Krug4}, a description of dynamical
interface fluctuations in terms of persistence and survival probabilities 
is not just an equivalent alternative to the usual 
description~\cite{book1,krugrev,book3} 
based on dynamical scaling:
first-passage properties provide a richer characterization
of the fully nonlocal and non-Markovian (both in space and time) nature of
interface fluctuations. This is reflected in the nontrivial behavior of
some of the first-passage properties even for fluctuation processes 
described by linear dynamical equations for which
dynamical scaling is rather trivial~\cite{Krug4}.

Our studies are also
motivated by the importance of step fluctuations in the practical
problem of assessing
the stability of nanoscale structures. This 
is a fundamental and challenging issue that will become
increasingly important as the length scale of devices approaches the atomic
limit~\cite{nature,science,science2}. 
In this context, it is important to address questions such as
what is the distribution of the time for a nanoscale structure to fluctuate
{\it for the first time} to a point of encounter at which a change 
in properties
or some other switching behavior may occur. Studies of first-passage 
properties of fluctuating steps and similar 
atomic-scale structures on surfaces 
are relevant for answering such questions.       

The rest of this paper is organized as follows. Various persistence and
survival properties studied by us are defined in section~\ref{def}.
The models used in our analytic and numerical studies are defined in
section~\ref{models}. The methods used in our numerical studies are also
summarized in this section. Section~\ref{expt} contains a brief account of
our experimental methods. The results of our analytic, numerical and
experimental studies of first-passage properties of equilibrium fluctuations
of isolated steps with different microscopic kinetics are described in detail
in section~\ref{results}.
Section~\ref{summ} contains a summary of our main results  and a 
few concluding remarks.

\section{Persistence and survival probabilities}\label{def}

We consider an isolated 1d step whose position at time $t$
is described by the function $h(x,t)$ where the $x$-axis is taken to 
be along the average step position. In temporal persistence and survival,
we consider the temporal fluctuations of $h(x,t)$ at a fixed position $x$.
The temporal persistence probability $P(x; t_0, t_0+t)$ is defined as the
probability that the sign of $[h(x,t_0+t^\prime)-h(x,t_0)]$ {\it does not
change} for {\it all} $0 < t^\prime \le t$. Clearly, this is the probability
that the step-edge position does not return to its initial value (at 
time $t_0$) during its stochastic time evolution over the period from $t_0$
to $t_0+t$. Averaging $P(x; t_0, t_0+t)$ over the coordinate $x$ and 
the initial time $t_0$ in the equilibrium 
state of the step, one obtains the temporal persistence probability $P(t)$
considered in our work. It is obvious that $P(t)$ is closely related to the
zero-crossing statistics of the stochastic variable $[h(x,t_0+t)-h(x,t_0)]$. 
In particular, if $w(t)$ is the distribution of the time interval between
two successive zero crossings of this quantity, then $P(t)=\int_t^\infty 
w(t^\prime) dt^\prime$.

The probability of {\it persistent large deviations}, introduced by Dornic
and Godreche~\cite{Dornic1}, provides a more detailed characterization of
stochastic processes.
A closely related idea, that of sign-time distribution, was developed in
Ref.~\cite{zoltan}. This involves the variable $r(t) \equiv
\hbox{sign}[h(x,t_0+t)-h(x,t_0)]$ that takes the values $\pm 1$. The average
sign variable $r_{av}(t)$ is defined as $r_{av}(t) 
\equiv t^{-1}
\int_{0}^{t} r(t^\prime) dt^\prime$. Clearly, the values of $r_{av}(t)$
lie between $-1$ and $1$. Then, the probability of 
persistent large deviations, $P(t,s)$, is defined as the probability
for the average sign $r_{av}$ 
to remain above a certain pre-assigned value $s$ 
($-1 \leq s \leq 1$), up to time $t$:
$P(t,s) \equiv \hbox{Prob}~\lbrace ~r_{av}(t^{\prime})
\geq s,~\forall 0< t^{\prime} \leq t~ \rbrace$. Averages over the initial time
$t_0$ and the coordinate $x$ are implied in the above definition. Also, we
assume, without any loss of generality, that the initial
deviation of the step position is in the ``positive'' direction, i.e. 
$h(x,t_0+t) > h(x,t_0)$ for $t \to 0$. Then, the initial value of 
$r_{av}(t)$ is equal to unity
and $P(t,s=1)$ is identical to the temporal persistence probability $P(t)$
defined above. Also, $P(t,s=-1)$ is trivially equal to unity for all $t$.
The time-dependence of $P(t,s)$ for $-1 < s < 1$ is quite nontrivial and
its dependence on $s$ provides a convenient way of characterizing the 
underlying stochastic dynamics.

The temporal survival probability $S(x; t_0, t_0+t)$ is defined as the 
probability that the sign of $[h(x,t_0+t^\prime)-\bar{h})]$ {\it does not
change} for {\it all} $0 < t^\prime \le t$. Here, 
$\bar{h}$ is the {\it thermodynamic average}
value of the step position which does not depend on the coordinate $x$. Thus,
the temporal survival probability $S(t)$, obtained by averaging 
$S(x; t_0, t_0+t)$ over $t_0$ and $x$, measures the probability of the step
position not crossing its average value during its evolution over time $t$.
This quantity is clearly related to the zero crossing statistics of the 
deviation of the step position from its average value. Although the definition
of $S(t)$ is rather similar to that of the persistence probability $P(t)$,
the time-dependences of these two quantities turn out to be quite different.

We have also considered a generalization of the  
survival probability
in which the ``return position'' is shifted from the average step
position $\bar{h}$ to an arbitrary reference position $R$. The generalized
(temporal) survival probability $S(t,R)$ is defined as the probability that
the step position $h$ is initially beyond the pre-assigned reference position,
$h>R$ (or $h <-R$, for fluctuations that are symmetric about the average 
value, $\bar{h}=0$), and remains there over a given time interval $t$. 
Clearly, $S(t,R=\bar{h})=S(t)$, but the time-dependence of $S(t,R)$ for
other values of $R$ depends on the choice of $R$. Another generalization,
the generalized inside survival probability $S_{in}(t,R)$, is
defined as the probability of the step position remaining between the 
pre-assigned reference positions $R$ and $-R$ (assuming the fluctuations to 
be symmetric about the average value, $\bar{h}=0$) over time $t$.

Spatial persistence and survival probabilities are defined in the same way as
the temporal
quantities mentioned above, but considering the step position $h(x,t)$ to be
a function of the coordinate $x$ for fixed $t$. For example, the spatial
persistence probability $P(x)$ is defined as the probability that the 
step position at a fixed time $t$ 
does not return to its ``initial'' value $h(x_0,t)$ as one
moves from the point $x_0$ to the point $x_0+x$ along the 
average step direction (averages over
$x_0$ and $t$ are implied).   

\section{Models and numerical methods} \label{models}

If the separation between adjacent steps is sufficiently large, the entropic
and elastic interactions~\cite{ellen,giesen} between different steps may be
neglected. We consider here this simple situation where the dynamics of a 
step is not affected by other steps on the surface. The energy of an isolated
step is then simply proportional to the total length measured along its
fluctuating edge. In a continuum description where the step-edge position is
denoted by the function $h(x,t)$, the Hamiltonian of an isolated step is then
given by
\begin{eqnarray}
\mathcal{H}[h(x)] &=& \tilde{\beta} 
\int_0^L [1+|\partial h/\partial x|^2]^{1/2} dx
\nonumber \\
&\simeq& \frac{\tilde{\beta}}{2} \int_0^L |\partial h/\partial x|^2 dx 
+\hbox{constant},
\label{hamiltonian}
\end{eqnarray}
where $\tilde{\beta}$ is the ``step-edge stiffness'' 
(energy per unit length), $L$ is the step size, and we have
used a small-gradient expansion. At relatively high temperatures, 
fluctuations in the step position 
are known~\cite{book1,book3,bartelt} to be dominated by random
attachment and detachment (AD) of atoms at the step edge. The ``noise'' arising
from these microscopic processes is clearly non-conserving: the total number of
atoms in the crystal layer that terminates at the step edge does not remain
constant when the AD mechanism is present. Under these
conditions, the dynamics of step fluctuations is described by the 
second-order non-conserved linear Langevin equation 
\begin{equation}
\frac {\partial h(x, t)} {\partial t} =
\frac{\Gamma_a \tilde{\beta}}{k_BT} \frac{\partial^{2} h(x,t)}
{\partial x^2} + \eta(x, t).
\label{eweqn}
\end{equation}
Here, $\Gamma_a$ is the ``step mobility'' and
and $\eta(x,t)$ is a non-conserved Gaussian noise
satisfying $\langle \eta(x, t) \eta(x^{\prime},t^{\prime})
\rangle= 2 \Gamma_a \delta(x-x^{\prime}) \delta(t-t^{\prime})$. This equation
is known in the literature as the Edwards-Wilkinson (EW) equation~\cite{EW}. 
In the
dynamics governed by this equation, the spatial average of the step position
exhibits a random walk in time. In order to eliminate spurious effects of
this random-walk behavior, we adopt the convention of measuring the step
position from its instantaneous spatial average. Thus, from now on, the
variable $h(x,t)$ will be used to 
denote the deviation of the step position at point $x$  
and time $t$ from its spatial average at that time. 

The AD mechanism freezes out at low
temperatures and the primary mechanism of fluctuations in this regime is the
step-edge diffusion (SED) of atoms~\cite{book1,book3,bartelt}. The 
noise is clearly conserving in this case: the integral of $h(x,t)$ over the
length of the step does not change as atoms diffuse along the step edge. 
Step edge fluctuations under these conditions are described by the 
fourth-order conserved
Langevin equation
\begin{equation}
\frac {\partial h(x, t)} {\partial t} =
-\frac{\Gamma_h \tilde{\beta}}{k_BT} \frac{\partial^{4} h(x,t)}
{\partial x^4} + \eta_c(x, t),
\label{conseqn}
\end{equation}
with $\langle \eta_{c}(x,t) \eta_{c}(x^{\prime},t^{\prime}) \rangle =
-2 \Gamma_h \nabla_x^2 \delta(x-x^{\prime})
\delta(t-t^{\prime})$. It is easy to check that Eqs.(\ref{eweqn}) and 
(\ref{conseqn}) lead to the same equilibrium behavior at long times, but due
to the conserved nature of the noise, the dynamics governed by 
Eq.(\ref{conseqn}) strictly conserves the integral of $h(x,t)$ over $x$.
The deterministic part of the Langevin equation (\ref{conseqn}) is the same
as that in the so-called Mullins-Herring equation for surface growth~\cite{MH},
but the noise in the Mullins-Herring equation is non-conserving. In a general
situation, both AD and SED may be present.
However, theoretical studies~\cite{th1,th2} suggest that a
single mass transport mechanism will be dominant 
for most physical systems. Therefore, we
can make the simplifying assumption of considering each mechanism separately,
and study the dynamics governed by one of the two
Langevin equations, (\ref{eweqn}) and (\ref{conseqn}). 

All space and time-dependent correlation functions for these two linear
Langevin equations can be calculated analytically. The equilibrium distribution
of $h(x,t)$ is Gaussian with zero mean and 
width equal to the root-mean-square width of the step, $W(L) \propto
L^\alpha$, with $\alpha=1/2$. The mean-square difference in the step
positions at two different points along the step at the same time is
given by
\begin{equation}
G(x) \equiv  \langle [h(x+x_0,t) - h(x_0,t)]^2 \rangle
\propto x^{2\alpha},
\label{gx}
\end{equation}
at equilibrium for $x \ll L$. Note that this dependence on $x$ is the same as 
the dependence of the mean-square displacement on time in a 1d
random walk. In fact, the spatial statistics of $h$ in the equilibrium state
is identical to the temporal statistics of the displacement in Brownian motion
in one dimension.

Although the two Langevin equations considered here lead to the same 
equilibrium behavior, time-dependent correlations are different in the two
cases. The equilibrium temporal correlation function of height fluctuations at
the same point decays exponentially at long times,
\begin{equation}
C(t) \equiv \langle h(x,t_0+t) h(x,t_0) \rangle \propto \exp[-t/\tau_c(L)],
\label{ct} 
\end{equation}
where the correlation time $\tau_c(L)$ is proportional to $L^z$, with $z=2$
for the EW equation and $z=4$ for the conserved fourth-order equation. Also,
the mean-square difference in the step
positions at the same point at two different times is
given by
\begin{equation}
G(t) \equiv \langle [h(x,t_0+t) - h(x,t_0)]^2 \rangle \propto t^{2 \beta},
\label{gt}
\end{equation}
for $t \ll \tau_c$, with $\beta = \alpha/z$, so that $\beta=1/4$ for the EW
equation and $\beta=1/8$ for the conserved fourth-order equation. 

Although the step fluctuation $h(x,t)$ is a Gaussian stochastic variable, its
non-Markovian nature makes analytic calculations of some of its first-passage
properties difficult. In our studies, we used numerical integrations of the 
Langevin equations (\ref{eweqn}) and (\ref{conseqn}) to evaluate some of 
their temporal and spatial first-passage properties. In these numerical
calculations, we used spatially discretized, dimensionless forms of the
Langevin equations which were integrated forward in time using a simple
Euler scheme. The dynamical equations considered in the 
numerical work are
\begin{equation}
\frac{dh_i}{dt} = (h_{i-1}+h_{i+1}-2h_i) + \eta_i(t), \label{ew}
\end{equation}
for the EW equation and
\begin{equation}
\frac{dh_i}{dt} = - (h_{i-2}-4h_{i-1}+6h_{i}-4h_{i+1}+h_{i+2}) +
\eta_i^\prime(t),
\label{4_th_order}
\end{equation}
for the conserved fourth-order equation. Here, $h_i(t)$ represents 
the step position at lattice site $i$ at
time $t$, 
$\eta_i(t)$ are uncorrelated random variables with zero mean and unit
variance, and 
and $\eta_i^\prime(t)$ are the conserved
version of such noise. Periodic boundary conditions were used in these 
calculations and results for different system sizes were obtained from 
averages over a large number of independent realizations of the stochastic
time evolution.

In some of our numerical studies, we also used Monte Carlo simulations of
the dynamics of 1d atomistic solid-on-solid models that are
known to belong in the same dynamical ``universality class'' 
as the two Langevin
equations mentioned above. For EW dynamics, we simulated the well-known 
Family model~\cite{F} in which
atoms deposited randomly, one at a 
time, at the latices sites are allowed to explore within a fixed
diffusion length to find the lattice site with the smallest ``height'' $h_j$ 
where it gets incorporated.
If the diffusion length is one lattice constant 
(this is the value used in our
simulations), the application of this deposition rule to a randomly
selected site $j$
involves finding the minimum value among the set 
$h_{j-1}, h_j$ and
$h_{j+1}$. The height of the site with the minimum height is
then increased by one. Since the average height continues to grow in time
in this model, the fluctuations of interest in the present context are obtained
by subtracting the instantaneous spatial average from the variables
$\{h_j\}$. 
An atomistic model proposed by 
Racz {\it et al.}~\cite{racz} provides a discrete
realization of the conserved dynamics of Eq.(\ref{conseqn}). 
In this model, the
nearest-neighbor height differences are restricted to $|h_{j+1}-h_j| \leq 2$.
In one simulation step, a site $j$ is randomly chosen. 
A diffusion move to a randomly
chosen neighbor takes place if the above restrictive condition 
is satisfied after the move; 
otherwise
a new random site is selected and the procedure follows in the same way. In
these atomistic simulations, ``time'' is measured in units of attempted
depositions or moves per lattice site.

\section{Experimental methods} \label{expt}

	The experimental results reviewed here were all extracted from 
variable-temperature scanning tunneling microscope (VT-STM) measurements of 
monatomic steps on solid surfaces.  In this section, experimental concerns 
generic to all STM-based measurements of 
first-passage 
statistics are addressed.  
More specific details such as surface preparation and 
characterization can be found in Refs.~\cite{Dan,alex,PTS,Dan2}. 
   
First passage measurements for surface steps have been performed under 
ultra-high vacuum conditions to allow for the preparation 
of surfaces with 
well-defined and reproducible structure and chemical composition.  
This level of 
control is crucial to the interpretation of experimental results in terms of 
theoretical models of interface fluctuations and also 
allows the use of repeated 
measurements on identically-prepared samples to reduce fluctuations in 
first-passage statistics. An important experimental demonstration that the
surfaces are well-enough controlled is the measurement of the distribution
(either temporal or spatial) of displacements of the step-edge. For an
equilibrium system, the distribution must be Gaussian.

\begin{figure}
\begin{center}
\includegraphics[height=6cm,width=6cm]{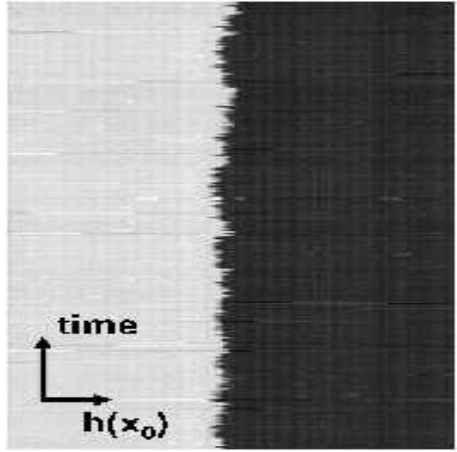}
\caption{\label{fig1} Example of line-scan pseudo-image of step 
fluctuations. In this figure, the step-edge position is denoted by $h$ and
$x_0$ denotes the coordinate in the average step direction.}
\end{center}
\end{figure}

	While STM is an obvious choice for real-space 
imaging experiments due 
to its remarkable spatial resolution, 
it is limited as a probe of dynamics by a 
relatively low data acquisition rate.  For most surfaces with step edges that 
fluctuate rapidly enough to generate useful temporal statistics, 
it is not possible 
to obtain an STM image that corresponds to a snapshot of step 
configurations at a 
single instant of time.  Therefore, the study of temporal 
step edge fluctuations 
often involves the use of so-called ``line-scan'' STM imaging.  
In this measurement, the STM tip is fixed at a point $x_0$ and scanned in one 
direction perpendicular to a step edge repeatedly for a 
fixed measurement time.  
An example of a resulting line-scan pseudo-image is shown in 
Fig.~\ref{fig1}
and represents a time series of the evolution of the step position at $x_0$.  
From this time series, $h(x_0, t)$, temporal first passage statistics can be 
extracted as described below.

	The low speed of the STM creates different difficulties for 
the measurement of spatial first-passage statistics 
where snapshots corresponding to a 
given instant of time are required.  To obtain these snapshots it is necessary 
to measure STM images at relatively low 
temperatures, where step fluctuations are 
slower than the rate of image acquisition, or to rapidly quench a surface from 
high temperature so that its step configuration is 
kinetically frozen~\cite{Dan2,lyubin}.  
Once obtained, spatial STM images provide the step 
configuration at fixed time, 
$h(x, t_0)$, from which first-passage statistics may be 
extracted as a function of $x$.  
Experimentally determined spatial first-passage statistics have 
been found~\cite{spexp} to be 
noisier than the analogous temporal quantities.  This arises partially due to 
the fact that, at large enough length scales along a step edge, there usually 
exist nonequilibrium features such as forced kinks due to small azimuthal 
crystal miscut or pinning sites due to trace contamination.  Therefore, the 
fraction of a spatial image that can be considered for quantitative analysis 
is usually smaller than the fraction of a temporal pseudo-image.

	The calculation of temporal persistence and survival 
from line-scan STM 
images proceeds differently than the calculation of these quantities 
from spatial 
images or from numerical simulations.  
In the latter two cases, probabilities are 
computed by calculating the fraction of all step sites that have not returned 
to the specified configuration (either as a function of 
time or of distance along 
the step).  For line-scan STM data that is obtained 
at a single point on the step 
edge, however, probabilities must be computed by calculating 
the fraction of all 
time intervals for which the given step position has not returned 
to the specified 
configuration.  For the measurements presented in the following sections, 
persistence and survival probabilities (as well as their generalizations) 
are computed from different STM images and averaged to 
obtain smooth curves that 
can be fit to theoretical predictions.  Parameters from these fits, such as 
persistence exponents and survival decay constants, are quoted as the average 
obtained from the different measurements and the experimental error in the 
parameter is taken as the standard deviation (one $\sigma$) 
of the different measurements.

\section{Analytic, numerical and experimental results} \label{results}

In this section, we present our results for various temporal and spatial
first-passage properties of equilibrium step fluctuations quantified in terms
of the persistence and survival probabilities defined in section~\ref{def}.
In each case, theoretical predictions (analytic when available and numerical)
are compared with the results obtained from STM-based experiments for 
systems with different microscopic mechanisms of mass transport along the
step edge.

\subsection{Temporal persistence probabilities} \label{tp}

For dynamics described by the linear Langevin equations (\ref{eweqn}) 
and (\ref{conseqn}),
the step-edge fluctuation $h(x,t)$ at a 
fixed position $x$ as a function of time
$t$ is a Gaussian stochastic process, 
but it is not Markovian. The non-Markovian
property arises from the presence of the spatial derivatives that 
generate ``interactions'' between height fluctuations at different space 
points. An exact analytic calculation of the persistence probability for 
a Gaussian but non-Markovian stochastic process is known~\cite{slepian}
to be very difficult. In fact, 
the persistence probability $P(t)$ can not be meanighfully defined
for the Langevin equations considered here 
if the time $t$ is considered to be a truly
continuous variable with no short-time cutoff.
As discussed in section~\ref{def},
$P(t)$ is defined in terms of the distribution of the time intervals
between successive zero-crossings of the stochastic variable
$Y(x,t) \equiv h(x,t_0+t) -h(x,t_0)$, with $t_0 \gg \tau_c \propto L^z$. 
Time-displaced correlation functions of this quantity, 
which  may be obtained from
Eq.(\ref{gt}), involve the exponent $\beta$. It can be shown~\cite{Krug1}
that for $\beta < 1$, which is the case for the Langevin equations considered 
here, the density of zero crossings is infinite -- once the process crosses
zero, it immediately crosses zero again many times before making a long 
excursion to the next zero crossing. In this case, the persistence
probability can be meaningfully defined only if a short-time 
cutoff is imposed. This, however, is not a serious problem
because Langevin equations such as the ones considered here are to be
understood as coarse-grained descriptions with naturally occurring short-time
and short-distance cutoffs. Also, in experiments and simulations, the 
step position is measured at discrete time intervals, so that the smallest
value of $t$ for which $P(t)$ needs to be defined is the sampling interval
$\delta t$. For this reason, there is no practical difficulty in measuring
$P(t)$ in experiments and simulations, although the mathematical problem
mentioned above implies that an exact analytic calculation of $P(t)$ for the 
models considered here is not possible.

It was shown in Ref.~\cite{Krug1} that for linear Langevin equation of the 
type being considered here, the persistence probability $P(t)$ decreases
in time as a power law, $P(t) \propto t^{-\theta}$, for $\tau_0 \ll t 
\ll \tau_c$ where $\tau_0$ is short-time cutoff. It was also shown that
the {\it persistence exponent} $\theta = 1-\beta$, where $\beta$ is the 
exponent defined in Eq.(\ref{gt}). This result was obtained from a scaling
argument and confirmed by simulations. The scaling argument is based on the
observation that the incremental autocorrelation function of the variable
$Y(x,t)$ defined above has a power-law form with exponent $\beta$,
\begin{equation}
\langle [Y(x,t_1)-Y(x,t_2)]^2\rangle \propto |t_1-t_2|^{2\beta},
\label{fbm}
\end{equation}
in the equilibrium state. This implies that the stochastic process $Y(x,t)$
is a {\it fractional Brownian motion}~\cite{FBM} with Hurst exponent $\beta$.
For such a process, it is known that the probability that, given it
has crossed zero at time $t=0$, it 
returns to zero after time
$\tau$ is proportional to $\tau^{-\beta}$, and
$N(T)$, the total number of 
zero-crossings up to time $T$ is proportional to $T^{1-\beta}$. Also, from the
definition of the persistence exponent $\theta$, it is clear that the 
number of zero-crossing intervals of length between
$\tau$ and $\tau +d\tau$
in the total interval $T$ is given by
\begin{equation}
n(\tau,T) \propto N(T) \tau^{-1-\theta} \propto T^{1-\beta} \tau^{-1-\theta}.
\end{equation}
Using the relation, 
\begin{equation} 
\int_0^T
d\tau\, \tau\, n(\tau,T)=T,
\end{equation} 
and equating powers of $T$ on both sides of this equation,
one then obtains
the result, $\theta = 1-\beta$. Simulations carried out in Ref.~\cite{Krug1}
and also by us have established the validity of this relation for the 
Langevin equations for step fluctuation.
Therefore, the persistence exponent should be 
$\theta = 3/4$ for Eq.(\ref{eweqn}) and $\theta = 7/8$ for Eq.(\ref{conseqn}).

\begin{figure}
\begin{center}
\includegraphics[height=8cm,width=8cm]{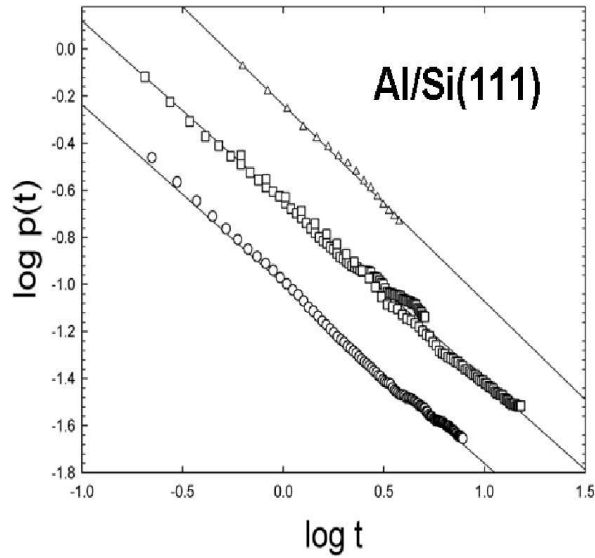}
\caption{\label{fig2} Double-logarithmic plots of the experimentally
obtained persistence probability $p(t)$ for Al/Si(111) surface steps, as a 
function of time $t$, for three different temperatures (770K, 970K and
870K, from top to bottom). The plots have been 
offset vertically from one another for clarity of display 
(from Ref.\cite{Dan}).} 
\end{center}
\end{figure}

The first experimental study of temporal persistence probability
for surface steps was carried out using measurements of step 
fluctuations on a
vicinal surface of Si(111)  modified by the adsorption of  Al~\cite{Dan}.
Previous experiments~\cite{lyubin,dan3} lead to the conclusion
that the fluctuations in this system
arise from the random exchange of mass between the step and the terrace
and that the interface should therefore be modeled using the EW
equation. Experimental persistence probabilities measured at three different
temperatures for the Al/Si(111) surface steps are shown in Fig.~\ref{fig2}.
It is apparent from the linear behavior of the double-logarithmic plots that
the persistence probability scales with time as a power law.  The average
persistence exponent for the Al/Si(111) surface is $0.77 \pm 0.03$, in good
agreement with the prediction of $\theta = 3/4$ 
for the EW equation.

Temporal persistence probabilities have also been extracted for
metal surfaces of Pb(111) and Ag(111) where steps are known to fluctuate due
to the SED mechanism~\cite{giesen,dan6,dan7}.  
The persistence probability has been found~\cite{alex} to decay in time as a 
power law in these cases too. 
For steps on
Pb(111), the persistence exponent is $\theta = 0.88 \pm 0.04$ and 
for steps on Ag(111),
the persistence exponent is $\theta = 0.87 \pm 0.02$.  
These exponents are in agreement
with the value of 7/8 predicted for the conserved fourth-order
Langevin equation (\ref{conseqn}).

The most important role of these experimental results is to confirm
the theoretical prediction of power law scaling of temporal persistence
with an exponent that depends on the mass transport mechanism governing the
relaxation of fluctuations.
Furthermore, the experimental results quantitatively
support the theoretically obtained relationship, $\theta=1-\beta$, between the
persistence exponent $\theta$
and the more commonly measured growth exponent $\beta$.
The persistence probability
can therefore be used as an independent means of establishing
the most appropriate
model to describe a fluctuating interface.

\begin{figure}
\begin{center}
\includegraphics[height=14cm,width=8cm]{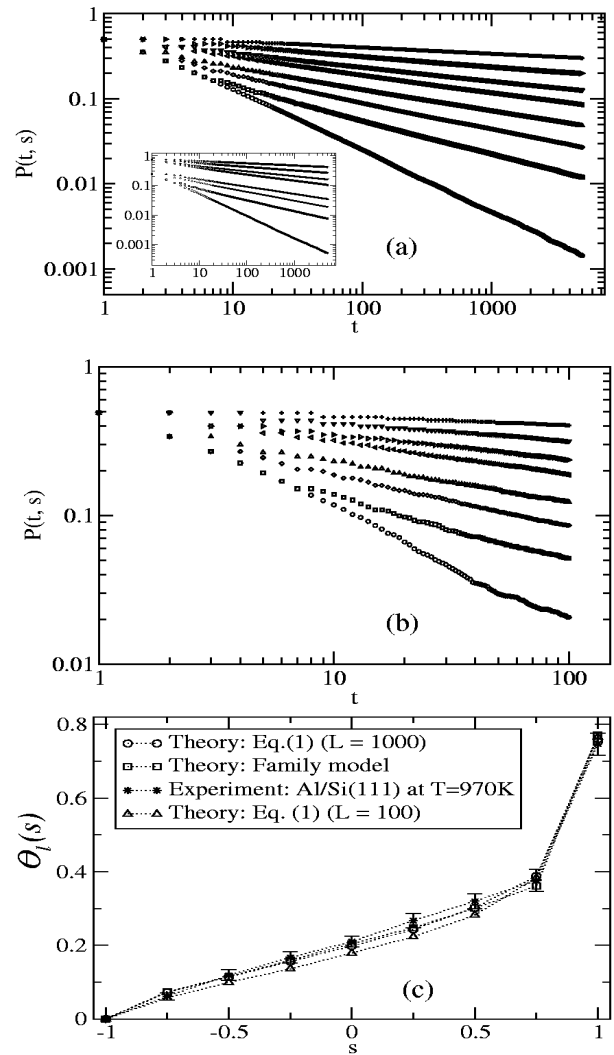}
\caption{\label{fig3} Log-log plots of $P(t,s)$ vs.~$t$ for
high-temperature surface step fluctuations via the AD
mechanism,
shown for different values of $s$:
$s=1.0,0.75,0.5,0.25,0,-0.25,-0.5,-0.75$ (from the bottom
to the top). (a) Eq.(\ref{eweqn}) (main figure) and the Family model
(inset); (b) experimental data from STM step images of Al/Si(111) 
surface at 970K;
and (c) comparison of the various sets of results for $\theta_l$
as a function of $s$. The error bars shown for the
experimental data are obtained from variations of the local
slope of the $\log P(t,s)$ vs.~$\log t$ plots. Simulation results
for two sample sizes are shown to illustrate that the use of
small samples leads to an underestimation of $\theta_l(s)$
(from Ref.~\cite{PTS}).}
\end{center}
\end{figure}

We have also investigated the behavior of
the probability of persistent large deviations $P(t,s)$, discussed in 
section~\ref{def}, for step fluctuations. 
It was found in Ref.~\cite{Dornic1} that for a number of
stochastic systems, the probability $P(t,s)$ decays in time as a power law,
$P(t,s) \propto t^{-\theta_l(s)}$, with the exponent $\theta_l(s)$ varying
continuously with $s$ from $\theta_l(s)=\theta$, 
the usual temporal persistence exponent,
for $s=1$ to $\theta_l(s)=0$ for $s=-1$. The {\it function} $\theta_l(s)$,
$-1 \leq s \leq 1$, therefore, defines an infinite family of continuously
varying exponents that characterizes the probability of persistent large 
deviations for the stochastic dynamical system under study. 
Analytic results for this function are available
only for a class of simple spin models~\cite{th_Pts} in which the 
intervals between successive spin-flips are assumed to be uncorrelated.
This assumption is not valid in the models of step fluctuation considered
here. We have investigated~\cite{PTS} the behavior of $P(t,s)$ numerically and 
and experimentally for step fluctuations with different mechanisms of 
microscopic mass transport. 

For step fluctuations in a regime dominated by the AD
mechanism, we have performed~\cite{PTS} numerical studies of the EW equation, 
Eq.(\ref{eweqn}), and the 1d Family model, and experiments on 
Al/Si(111) steps.
In Fig.~\ref{fig3} we have shown three sets of results
for $P(t,s)$, obtained from numerical integration of the EW equation,
discrete stochastic simulation of the Family model, 
and analysis of the experimental data, respectively.
It is clear that the decay of $P(t,s)$ at equilibrium follows a power law,
$P(t,s)\propto t^{-\theta_l(s)}$, characterized by a persistent large
deviation exponent $\theta_l(s)$ that varies continuously between $0$
(for the limiting case $s=-1$) and the steady--state persistence exponent
$\theta=3/4$ (for the $s=1$ case) discussed in the first part of this section.
All three sets of results for the persistent large deviations exponents 
agree very well, establishing the infinite family of persistence
exponents as a potentially powerful tool in studying dynamical
interface fluctuation processes.

A similar set of exponents for persistent large deviations was also
determined~\cite{PTS} from numerical studies of 
Eq.(\ref{conseqn}) and the atomistic model of Racz {\it et al.}
described in section~\ref{models}, and from Ag(111) step fluctuation 
measurements. The qualitative shape of the dependence of the experimentally
obtained persistent large deviations
exponent on the reference level $s$ agrees well with the numerical results,
but the quantitative agreement between numerical
and experimental results in this case is not as good as it is for Al/Si(111).
The reason for this discrepancy is not fully understood at present.
There are some indications that the limited time range of 
experimental data
or incomplete equilibration of numerical models might play 
a detrimental role to
the correspondence between theory and experiment.
A more intriguing possibility is that some aspects of mass transport 
on Ag(111)
are not captured by the simplest form of the SED mechanism.
This situation illustrates the potential value of investigating 
a broad range of
statistical quantities, since all 
possible subtleties in fluctuation mechanisms
may not be apparent from equilibrium correlation functions 
or even standard
persistence probabilities.

\subsection {Temporal survival probabilities} \label{ts}

For step dynamics described by the linear Langevin equations (\ref{eweqn})
and (\ref{conseqn}), the equilibrium 
time-autocorrelation function of the fluctuations
of the step position decays exponentially at long times (see Eq.(\ref{ct})). 
This implies that $h(x,t)$ is a stationary Gaussian Process with exponentially
decaying autocorrelation function. A well-known result~\cite{pers} in the
theory of stochastic processes then implies that the temporal survival
probability $S(t)$ that measures the probability of $h(x,t)$ not crossing
its average value $\bar{h} =0$ over time $t$ should also decay exponentially
in time with a time constant $\tau_s$ that is proportional to the correlation
time $\tau_c$. The constant of proportionality $c=\tau_s/\tau_c$ is  known
to be less than unity and independent of the system size $L$, but its value
is non-universal, being determined by the full functional form of the
autocorrelation function $C(t)$. Thus, we arrive at the interesting result 
that although the definitions of the persistence probability $P(t)$ and the
survival probability $S(t)$ appear to be 
quite similar, the fact that the reference
levels used in their definitions are different (the initial step position
for $P(t)$ and the average step position for $S(t)$) results in a 
qualitative difference in the long-time behaviors of these two quantities.
Persistence probabilities convey, through the values of the persistence
exponent $\theta$ and the infinite family of exponents $\theta_l(s)$ for 
persistent large deviations, information about the dynamical universality
class of step fluctuations, whereas the survival probability conveys 
information about the long-time relaxation 
of step fluctuations.
It should be noted that this difference arises from the continuous nature of
the variable $h(x,t)$. Persistence and survival probabilities would be 
identical in problems involving discrete variables such as Ising spins for 
which a flip of the sign of a spin ensures a change of sign with respect to 
both initial and average values of the stochastic variable. 

\begin{figure}
\begin{center}
\includegraphics[height=11cm,width=15cm]{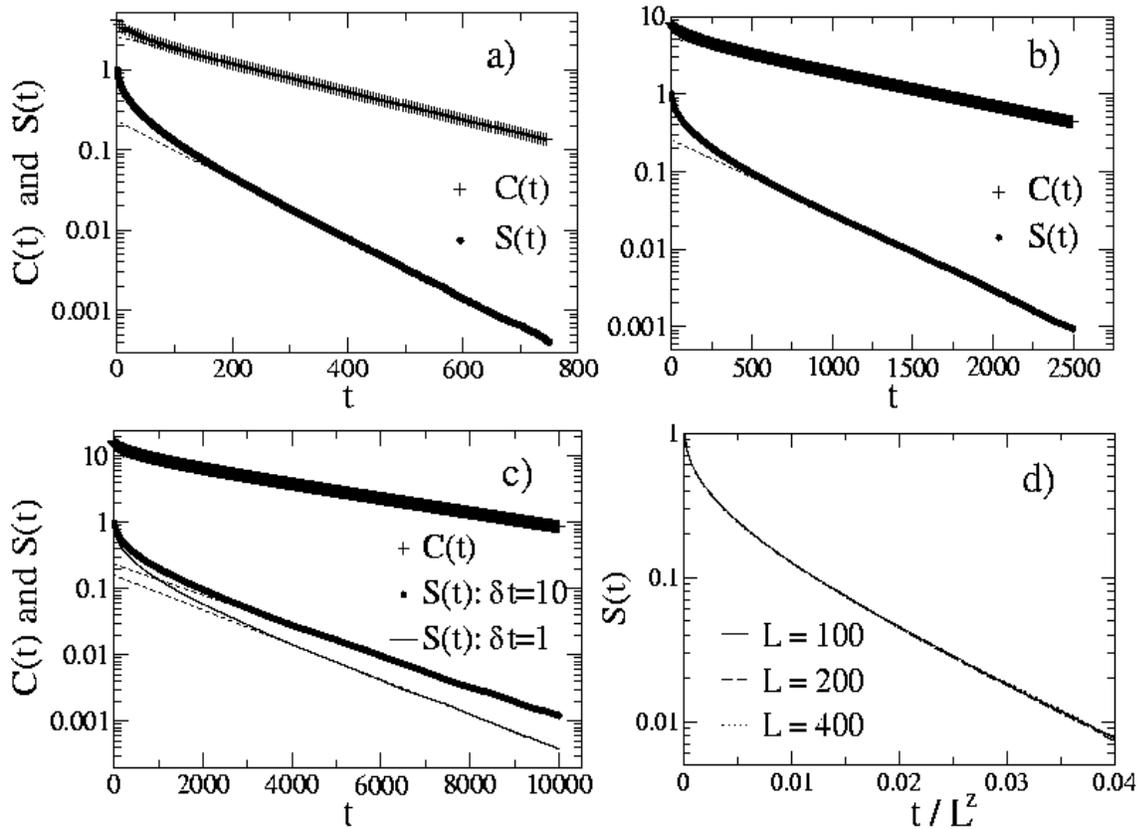}
\caption{\label{fig4} The survival probability 
$S(t)$ and the autocorrelation function $C(t)$ for the Langevin equation of
Eq.(\ref{eweqn}). The dashed lines are fits
of the long-time data to an
exponential form. In panels (a-c), the uppermost plots show
the data for $C(t)$. Panel (a): $L=100$, $\delta t=0.625$. Panel (b):
$L=200$, $\delta t=2.5$. Panel (c): $L=400$, $\delta t=10.0$
(upper plot) and $\delta t=1.0$ (lower plot).
Panel (d): Finite-size scaling of $S(t,L,\delta t)$.
Results for $S$ for 3 different sample sizes with
the same value of $\delta t/L^z$ ($z=2$) are plotted 
versus $t/L^z$ (from Ref.~\cite{ST}).}
\end{center}
\end{figure}

The exponential decay of $S(t)$ has been confirmed from simulations~\cite{ST}
and experiments~\cite{Dan,Dan2,expt3} for both AD 
and SED dominated kinetics. Typical simulations results 
for $C(t)$ and $S(t)$ are shown in Fig.~\ref{fig4}
(a-c) for dynamics governed by the EW equation.
While the $L$-dependence of $\tau_c$ is known
exactly, our results for $S(t)$ reveal the fact that although $\tau_s$
increases rapidly with $L$, as expected, there are clear deviations
from the expected proportionality to $L^z$ if {\it the sampling
time $\delta t$} (i.e., the time between consecutive measurements of the 
step-edge position)
is kept unchanged and only $L$ is varied. Moreover, the scaling
behavior of $S(t)$ can be revealed only if $\delta t$ is
carefully incorporated into the analysis. We explain in detail the important
role played by the sampling time on the persistence and survival probabilities
in the next subsection. Similar results were also obtained~\cite{ST} for the
fourth-order conserved Langevin equation, Eq.(\ref{conseqn}).

\begin{figure}
\begin{center}
\includegraphics[height=6cm,width=12cm]{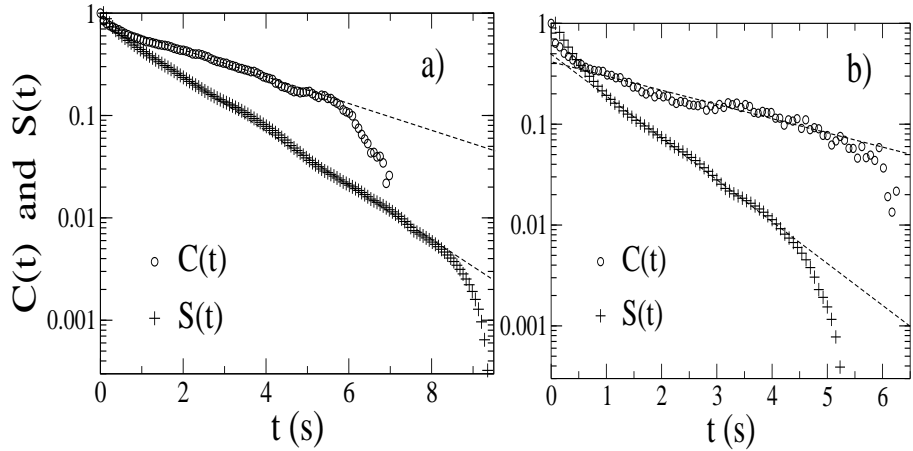}
\caption{\label{fig5} $S(t)$ and $C(t)$ for two experimental
systems. The dashed lines are fits of the long-time data to
an exponential form. Panel (a): Al/Si(111) at $T$ = 970K.
Panel (b): Ag(111) at $T$ = 450K (from Ref.~\cite{ST}).}
\end{center}
\end{figure}

In Fig.\ref{fig5}, the experimental survival probabilities and autocorrelation
functions~\cite{Dan,ST} are shown for Al/Si(111) and Ag(111)
on a semi-logarithmic plot.
For both material systems, exponential decay is observed for temporal survival
and autocorrelation even though the fluctuations on Al/Si(111) belong to a
different model universality class than those on Ag(111).
The detailed quantitative analysis
of the survival probability depends on experimental subtleties arising
from discrete sampling and finite measurement time. These issues will be 
discussed in the next subsection.

As noted in section~\ref{def}, the 
temporal survival probability can be generalized~\cite{STS}
by considering probabilities
associated with the interface position not crossing an arbitrarily chosen
reference level $R$, as opposed to the equilibrium average position used in
the definition of $S(t)$.
This so-called generalized survival probability $S(t,R)$ 
is a natural means of addressing
stability issues in nanostructured environments
where the significant reference
level is unlikely to be the precise equilibrium average interface position.
For equilibrium step fluctuations, the generalized survival
probability $S(t,R)$ is defined 
as the probability for the step position to remain consistently
{\it above} a certain pre-assigned value $R$ (or, equivalently, {\it below}
$-R$, since the step fluctuations are symmetric about zero) over time $t$.

\begin{figure}
\begin{center}
\includegraphics[height=8.0cm,width=10.0cm]{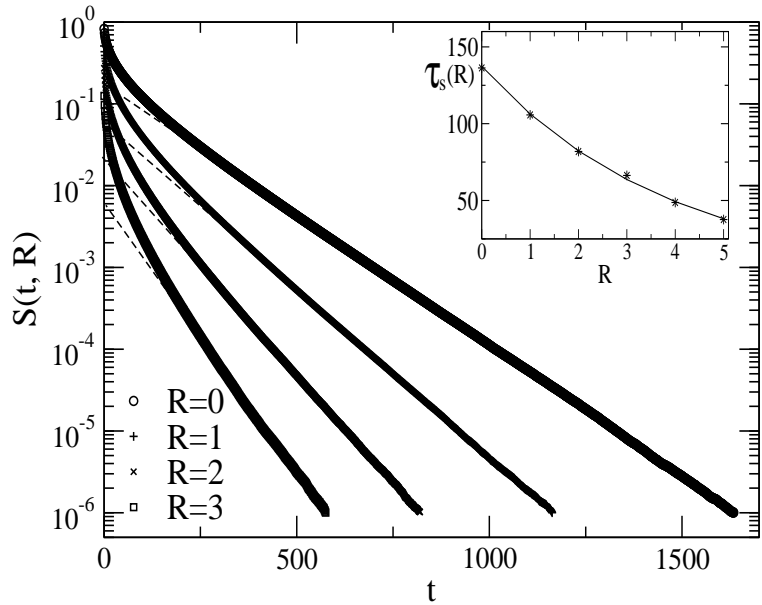}
\caption{\label{fig6} $S(t,R)$ for the discrete Family model.
The dashed lines are fits of the long--time data to an exponential
form. The system size is $L=100$, the sampling time is
$\delta t=1.0$ and the reference level $R$ takes four
different values: $0$, $1$, $2$ and $3$ (from top to
bottom). The inset shows the dependence of the generalized survival
time scale $\tau_s(R)$ on the reference level value (up to $R=5$).
The continuous curve represents a fit to an exponential decay of
$\tau_s(R)$ with $R$ (from Ref.~\cite{STS}). }
\end{center}
\end{figure}

Numerical results~\cite{STS} for the generalized survival probability and the
associated time scale, obtained from simulations of the 1d
Family model, are presented in Fig.~\ref{fig6}. For all $R$, $S(t,R)$
decays exponentially in the long-time limit, with an associated
time scale, $\tau_s(R)$, which decreases with the reference level
value. As shown in the inset of Fig.~\ref{fig6}, the dependence of
$\tau_s(R)$ on $R$ appears to be exponential. At present, no
analytic results are available for the time-dependence of $S(t,R)$ in the
models being considered here. 

\begin{figure}
\begin{center}
\includegraphics[height=12.0cm,width=10.0cm]{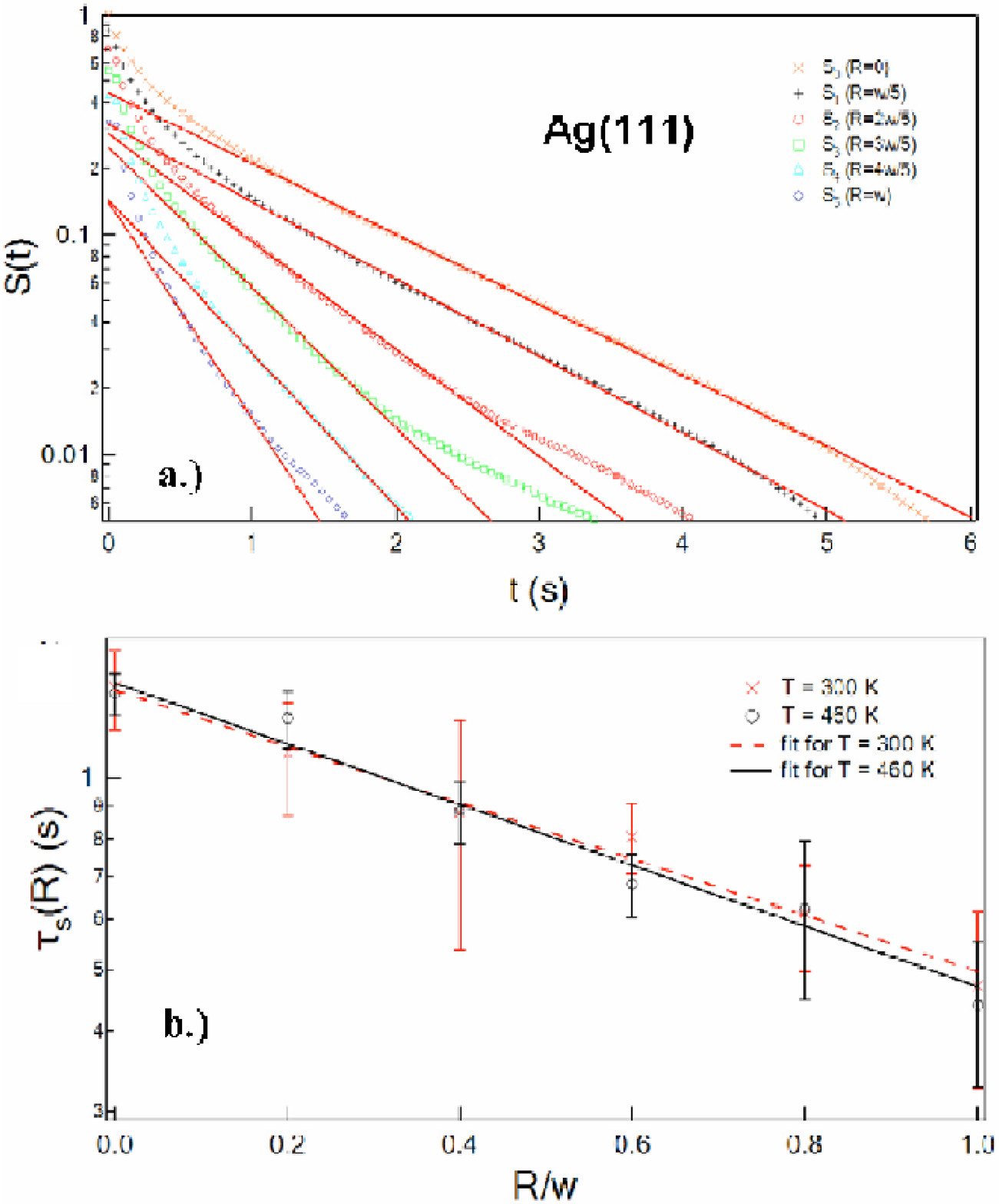}
\caption{\label{fig7}
Experimental data (top panel) for the generalized survival
probability $S(t,R)$ for Ag(111) steps. The bottom panel shows the variation
of the time constant $\tau_s(R)$ with the scaling variable $R/W$ where
$W$ is the measured root-mean-square fluctuation 
of the step position (from Ref.~\cite{unpub}).}
\end{center}
\end{figure}

Recently, the behavior of the  generalized survival probability
has been addressed~\cite{unpub}
experimentally for the case
of fluctuating steps on Ag(111) (kinetics governed by SED).
The exponential decay of $S(t,R)$
in time for all values of $R$ has been confirmed for this system, both
through STM measurements and
through numerical integration of the corresponding fourth-order 
conserved Langevin
equation. The experimental data shown in 
the top panel of Fig.~\ref{fig7} suggest a scaling
variable for the decay constant $\tau_s(R)$  in the form of the
ratio of the chosen reference level $R$ to the measured 
root-mean-square fluctuation
$W$ of the step position.  
Empirical fits to the dependence of $\tau_s(R)$ on $R/W$, 
shown in the bottom panel of 
Fig.~\ref{fig7} by the dashed lines, suggest that the 
scaling function
has a simple exponential form, in agreement with the simulation results
for EW dynamics shown in Fig.\ref{fig6}.

Finally, the generalized inside
survival probability $S_{in}(t,R)$ defined in 
section~\ref{def} has also 
been studied numerically and experimentally~\cite{unpub} for step fluctuations
governed by SED.
Like the generalized survival probability,
this quantity is expected to be a valuable characterization
of stability and reliability
in environments consisting of closely-spaced
thermally fluctuating nanostructures. In both experiments and simulations,
$S_{in}(t,R)$ is found to decay exponentially in time for all
$R$. The time constant $\tau_{in}(R)$ of this decay scales with
$R/W$, but the scaling function is different from that found for the 
generalized survival probability. 
In particular, it is found that $\tau_{in}(R)$
increases with $R/W$, and the dependence is not well-described by an 
exponential function. The experimental results for the time constant are
found~\cite{unpub} to be 
in good agreement with numerical predictions  when the effects of discrete
sampling and finite observation time, discussed in detail in the
next subsection, are taken into account.

\subsection{Effects of discrete sampling, finite system size and finite
observation time} \label{finite}

As noted above, the step position
$h(x,t)$ is measured in experiments and simulations at discrete intervals 
of a sampling time $\delta t$. 
It has been pointed out in Ref.~\cite{deltat,samp} that
discrete--time sampling of a continuous--time stochastic process
does affect the measured persistence and survival probabilities. 
Increasing the sampling
interval increases the persistence and survival probabilities because
some of the crossing events detected in sampling with a small $\delta t$
are missed if the step position is sampled with a larger $\delta t$. 
The results shown in Fig.~\ref{fig4}, panel
(c) illustrate this important fact -- the measured survival probability
exhibits appreciable 
dependence on the value of $\delta t$. 
Also, simulations are
always carried out for finite systems, and as explained below, the finiteness
of the time over which measurements are carried out translates into a finite
effective system size. Therefore, an understanding of
the effects of the finiteness of the sample size $L$ and the sampling
time $\delta t$ 
on the measured persistence and survival
probabilities and their generalizations
is crucial for extracting
reliable values of parameters such as 
the persistence exponents and survival time scales
from experiments and simulations.

We have used numerical simulations to examine in detail the effects of 
discrete sampling and finite system size on the measured first-passage
probabilities. We have found~\cite{ST,STS,PT} that the dependence of the
measured persistence and survival probabilities on the sampling time
$\delta t$ and the system size $L$ exhibits simple scaling behavior in terms
of the dimensionless scaling variables $t/\tau_c(L)$ and $\delta t/\tau_c(L)$.
This is reasonable, since the correlation time $\tau_c(L)$ is the only
relevant time scale in these systems.
The scaling behavior of the temporal survival probability is illustrated
in Fig.~\ref{fig4}, panel (d) where it is shown that plots of $S(t)$ 
versus the scaling variable $t/\tau_c(L)$ (recall that $\tau_c(L) 
\propto L^z$) for different $L$ and $\delta t$ all collapse to the same
scaling curve if $\delta t/L^z$ is held fixed. These results 
are for the 1d EW equation for which $z=2$. A similar
scaling behavior was also found~\cite{ST} for the survival probability
for the fourth-order conserved Langevin equation of Eq.(\ref{conseqn}).
From these results, we conclude that the dependence of the 
survival probability on the sampling interval and the sample size is
described by the 
following scaling form:
\begin{equation}
S(t,L,\delta t) = f_S(t/L^z,\delta t/L^z),
\label{scalingST}
\end{equation}
where the function $f_S(x,y)$ decays exponentially for large values of $x$ and
the rate of this decay increases slowly as $y$ is decreased.

\begin{figure}
\begin{center}
\includegraphics[height=6cm,width=15cm]{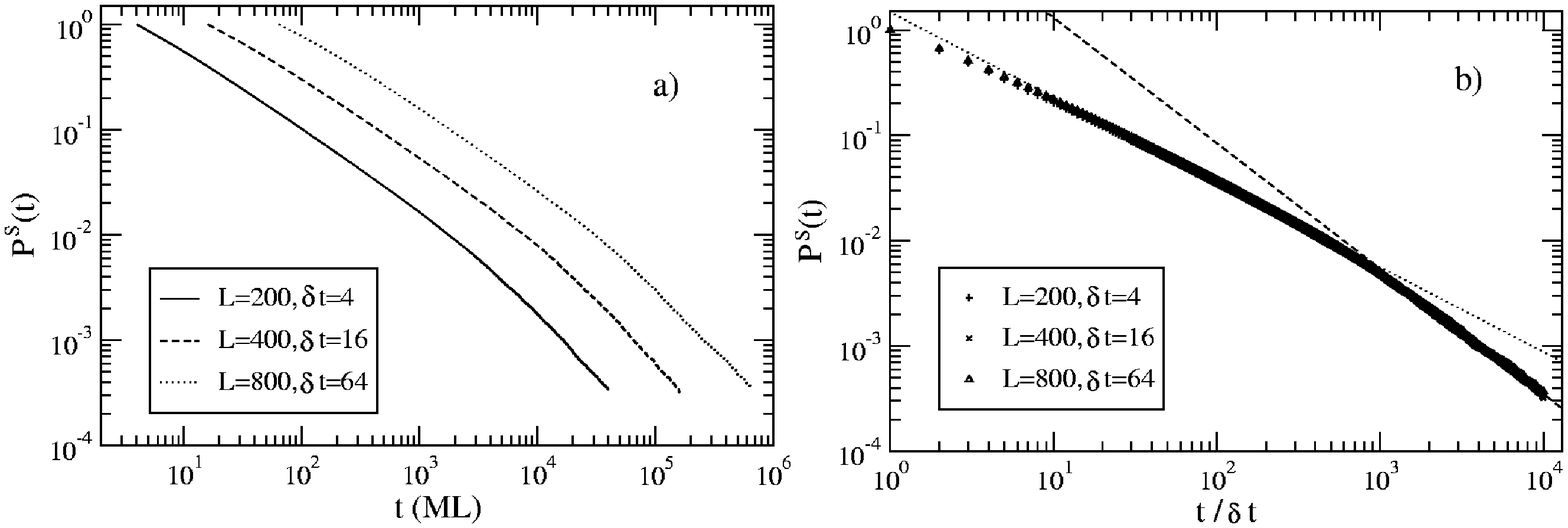}
\caption{\label{fig8}
Persistence probability, $P(t)$, for the Family model shown for
different system sizes with different sampling times.
Panel a): Double--log plot
showing three different $P(t)$ vs. time $t$ curves corresponding to:
$L=200$ and $\delta t=4$, $L=400$ and $\delta t=16$, $L=800$ and
$\delta t=64$, from top to bottom, 
respectively. Panel b): Finite size scaling of
$P(t,L,\delta t)$. Results for the persistence probabilities for three
different sizes (as in panel a) with the same value of $\delta t/L^z$
(i.e. $1/10^4$) are plotted versus $t/\delta t$ ($z=2$). The dotted (dashed)
line is a fit of the data to a power law with an exponent
$\simeq 0.75$ ($\simeq 1.0$) (from Ref.~\cite{PT}).}
\end{center}
\end{figure}

We have found
a similar scaling behavior for the steady--state persistence probabilities
measured in our simulations: 
\begin{equation}
P(t,L,\delta t)=f_P(t/L^z, \delta t/L^z),
\label{pscaling}
\end{equation}
where the scaling function $f_P(x,y)$ should decay as
$x^{-\theta}$ for small $x$ and $y \ll 1$.
In Fig.~\ref{fig8}, we show that the dependence of the steady--state
persistence probability in the Family model 
($z=2$) on $L$ and $\delta t$ is described by this scaling
form. Note that the scaling function $f_P$
exhibits the expected power-law behavior for relatively small values of
$t/\delta t$. However, we see signatures of a crossover
to a power-law decay with exponent 1 as $t$ approaches and
exceeds the characteristic time scale $\tau_c(L)\propto L^z$. This
behavior may be understood~\cite{PT} from the fact that height fluctuations
at two times separated by an interval that is much larger that $\tau_c$
are essentially uncorrelated. We have also shown~\cite{STS} that the 
generalized survival probability $S(t,R)$ exhibits a similar scaling
behavior. A non-zero value of the reference level $R$ introduces a new
length scale 
that is set by the equilibrium value of the interface
width $W \propto L^\alpha$. Therefore, the scaling
form of the generalized survival probability is expected to be
\begin{equation}
S(t,L,R,\delta t) = g(t/L^z,R/L^{\alpha},\delta t/L^z),
\label{scalingSTR}
\end{equation}
where the function $g(x,y,z)$ decays exponentially for large values
of $x$. The validity of this scaling form for the generalized survival
probability for the 1d Family model has been established
in Ref.\cite{STS}.
The use of the
scaling variable $R/W$ in the analysis of experimental data for 
generalized survival probabilities (see 
Fig.~\ref{fig7}) is motivated by this scaling relation. The observed
collapse of the data for $\tau_s(R)$ at two different temperatures to the
same curve when plotted as a function of $R/W$ (the values of $W$ at the
two temperatures are quite different) confirms the validity of the
theoretically predicted scaling behavior.

\begin{figure}
\begin{center}
\includegraphics[height=10cm,width=8cm]{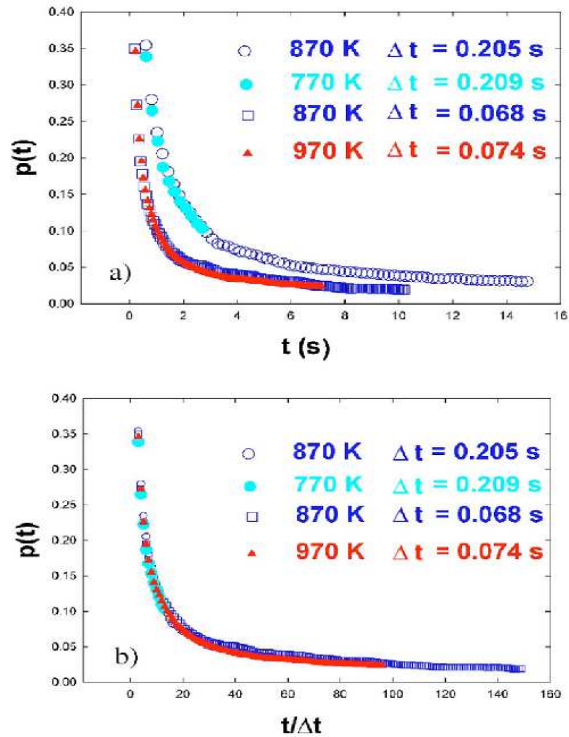}

\caption{\label{fig9}
Experimentally determined persistence probabilities $P(t)$
for steps on 
Al/Si(111) are shown in panel (a) as a function of time $t$
for different values of the temperature $T$ and the sampling interval
$\delta t$. The different curves collapse to the same one (panel (b)) when
the persistence probabilities are plotted versus $t/\delta t$
(from Ref.~\cite{Dan2}).}
\end{center}
\end{figure}

Several other experimental observations~\cite{Dan2,expt3,unpub} can be
understood in terms of the effects of finite sampling interval and system
size discussed above. Thermally-activated surface mass transport leads to 
strongly temperature-dependent
linear kinetic parameters that enter as coefficients in the Langevin equations
describing step motion~\cite{ellen,giesen,lyubin}. This 
temperature dependence is readily
apparent in experimentally-determined correlation functions.
Remarkably however, experimental persistence and survival probabilities
have not been observed to depend systematically on temperature. This
is due to the dependence of these quantities on the 
sampling interval and the (effective) sample size. The
experimental situation for the persistence probability is illustrated in
Fig.\ref{fig9} for Al/Si(111) step fluctuations~\cite{Dan2}. 
It is seen from
the data in panel (a) that when step positions are sampled 
with the same value of $\delta t$,
persistence probabilities are indistinguishable even at very 
different temperatures. 
In panel (b), it is shown that the persistence probabilities 
from 770 K to 970 K collapse to the same curve   
when plotted versus time scaled by the sampling interval $\delta t$
even though the underlying   
linear kinetic parameters vary by about two orders
of magnitude~\cite{giesen}. This can be understood from the scaling relation
of Eq.(\ref{pscaling}). Since the correlation time of the experimentally
studied surface steps is much larger than the observation time, 
Eq.(\ref{pscaling}) implies that the measured persistence probability
should depend only on the scaled time $t/\delta t$:  
in this regime, the sampling interval sets
the overall normalization of the persistence probability.     
It is important to point out that this discrete
sampling ``artifact'' does not affect the shape of the persistence 
probability (i.e the value of the
persistence exponent) but only its absolute magnitude. This behavior
is likely to be found in the measurement of
any first-passage probability using a
sampling rate slower than the underlying physical
processes and must be considered
carefully in analyzing experimental data.

In experimental studies of temporal survival probabilities, another
``artifact'' arises from the finiteness of the total measurement time 
$t_m$. This is because definitions of survival probabilities require the 
value of the average step position $\bar{h}$: the temporal survival 
probability measures the probability of the step position 
not returning to $\bar{h}$, and the reference level $R$ in measurements of
the generalized survival probability is defined relative to $\bar{h}$. In
analytic and numerical studies, $\bar{h}$ is known to be equal to zero.
In experiments, however, $\bar{h}$ is set to be equal to the average of
the step position measured over the time interval $t_m$. This average is
generally not the ``true'' average step position because fluctuation modes
with relaxation times much larger than $t_m$ do not equilibrate during the
observation time which is 
limited by the lateral stability of
the microscope. Analytic and numerical calculations described in 
Ref.~\cite{expt3} show that the main effect of the finiteness of $t_m$ is
very much similar to that of having a finite effective system size 
$L_{eff} \propto t_m^{1/z}$, with $z=2$ for AD
dominated kinetics and $z=4$ for SED limited kinetics.
Using this result (i.e. setting $L=L_{eff}(t_m)$) 
in the scaling relations (\ref{scalingST}) 
and (\ref{scalingSTR}) that describe the dependence of survival probabilities
on the system size, one obtains new scaling relations that describe the 
expected dependence of the experimentally measured survival probabilities
on the sampling interval $\delta t$ and the measurement time $t_m$.
These scaling relations involve the scaling variables $t/t_m$ and 
$\delta t/t_m$. The validity of these scaling relations has been 
established~\cite{expt3,unpub} from experimental measurements of the
temporal survival probability and the generalized survival probability
for different values of sampling time and observation time. For example,
it has been shown in Ref.~\cite{expt3} that for steps on Ag(111), plots
of the survival probability versus $t/\delta t$ for different values of 
$t_m$ and $\delta t$ collapse to the same scaling curve only when the
ratio $\delta t/t_m$ is held fixed. Since the effective system size is limited
by the observation time $t_m$, the time scales for the decay of the
measured autocorrelation function of step-edge fluctuations and the related
survival probabilities are all expected to be of order $t_m$. This explains
why the measured first-passage properties do not show any systematic
dependence on the temperature.

\subsection {Spatial persistence and survival probabilities}

Spatial counterparts of the temporal persistence and survival probabilities
may be defined by considering the equilibrium profile $h(x,t)$ of a step
at a fixed time $t$ as a function of $x$. 
The spatial persistence probability 
$P(x_{0},x_{0}+x)$ is simply the probability that
$h(x,t)$ {\it does not} return to its ``original''
value $h(x_0,t)$ at the initial point $x_0$ within a distance
$x$ measured from $x_{0}$ along the average step direction. In the 
statistically time-independent equilibrium state, this probability does not
depend on $t$.
A theoretical study~\cite{maj1} of this probability
for Gaussian interfaces with dynamics
described by linear Langevin equations shows that its dependence
on $x$ has a power-law form with an exponent that depends on how the initial
point $x_0$ is selected. If $x_0$ is sampled uniformly from {\it all} the
points of a steady-state interface configuration, then the average of 
$P(x_{0},x_{0}+x)$ over $x_0$ yields the
{\it steady-state} spatial persistence probability 
$P_{SS}(x)$ that decays with $x$
as a power law with an exponent $\theta_{SS}$ ($P_{SS}(x) 
\propto x^{-\theta_{SS}}$)
known as the steady-state spatial persistence exponent. If, on the other
hand, the initial point $x_0$ is sampled from a subset of points of a
steady-state configuration for which the values of $h(x,t)$ and its
spatial derivative are finite, then the so-called 
{\it finite-initial-conditions} spatial persistence probability is obtained,
which decays with $x$ as a power law with an exponent 
that may be different from
$\theta_{SS}$. The values of these exponents 
for interfaces with dynamics described by a class of linear Langevin
equations have been determined in Ref.~\cite{maj1} using a
mapping of the spatial statistical properties of the interface
to the temporal properties of
stochastic processes described by a generalized random-walk equation.
Steady-state and finite-initial-conditions spatial 
survival probabilities are
defined in a similar way in terms of the probability of the interface
not crossing its average value over distance $x$.    

We have carried out numerical studies~\cite{PX} of these spatial first-passage
probabilities for the models of equilibrium step fluctuation described in
section~\ref{models}. Both steady-state and finite-initial-conditions 
probabilities defined above were considered in these studies. For brevity,
we describe below the results obtained for the 
steady-state spatial persistence
and survival probabilities. Since the equilibrium spatial properties
are the same for the two
Langevin equations (\ref{eweqn}) and (\ref{conseqn}), we consider only the
EW equation and the Family model that belongs in the same universality 
class.

As discussed in section~\ref{models}, the incremental spatial autocorrelation
function of the variable $h(x,t)$ 
in the equilibrium state exhibits a power-law behavior with exponent
$\alpha=1/2$ (see Eq.(\ref{gx})). 
It is, therefore, clear that the incremental
spatial autocorrelation function 
of the variable $Z(x,t) \equiv h(x_0+x,t) - h(x_0,t)$,
whose zero-crossing statistics determines the spatial persistence probability
$P_{SS}(x)$, also exhibits a power-law behavior with the same exponent:
\begin{equation}
\langle [Z(x_1,t)-Z(x_2,t)]^2\rangle \propto |x_1-x_2|^{2\alpha}.
\label{fbm2}
\end{equation}
Comparing this with Eq.(\ref{fbm}) and using arguments identical to those
used for obtaining the relation between the temporal persistence exponent
$\theta$ and the dynamical exponent $\beta$, one readily obtains the result
that the steady-state spatial persistence exponent for EW interfaces is
given by $\theta_{SS} = 1-\alpha = 1/2$. This result also follows from the 
exact mapping~\cite{maj1} between the spatial 
statistics of a steady-state EW interface and
the temporal statistics of Brownian motion. 
This mapping
implies that the spatial persistence exponent $\theta_{SS}$ for a steady-state
EW interface is equal to the temporal persistence exponent, $\theta=1/2$,
for Brownian motion.

\begin{figure}
\begin{center}
\includegraphics[height=6cm,width=14cm]{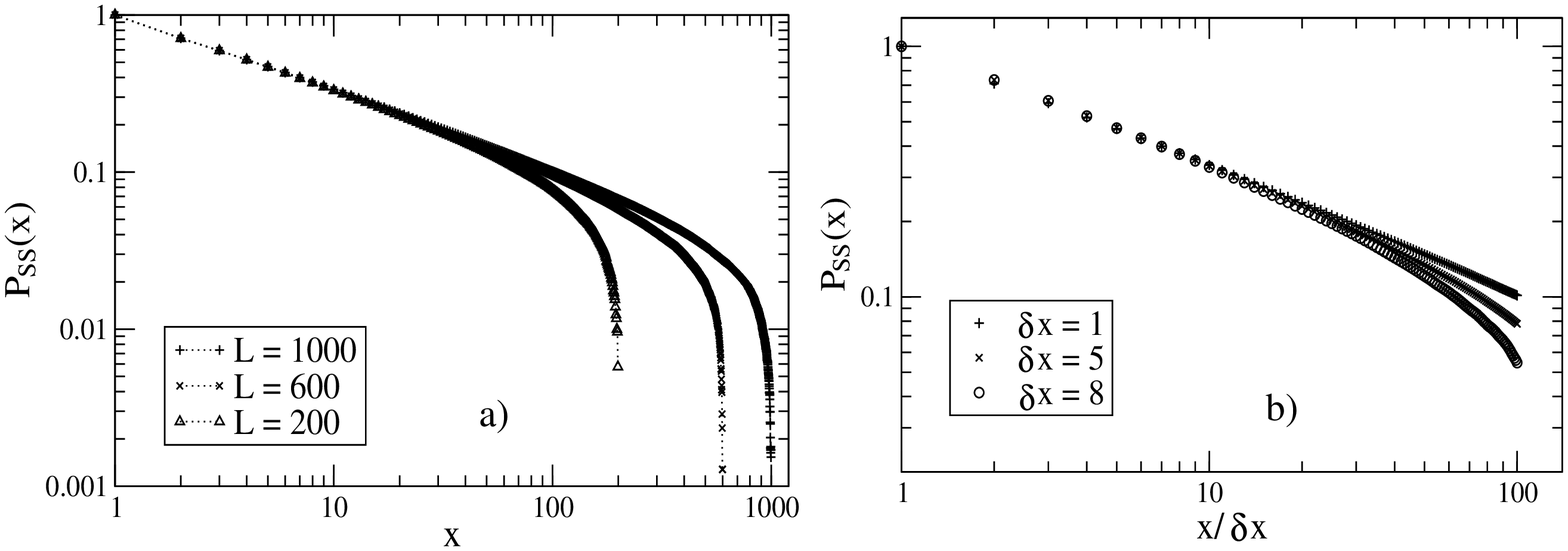}
\caption{\label{fig10} The steady state spatial persistence
probability, $P_{SS}(x)$, for EW interfaces, obtained 
using the discrete Family model.
Panel (a): Double-log plots of $P_{SS}(x)$ vs $x$ for a 
fixed sampling distance
$\delta x=1$, using three different values of $L$,
as indicated in the legend.
Panel (b): Double-log plots of $P_{SS}(x)$ vs $x/\delta x$ for a
fixed system size,
$L=1000$, and
three different values of $\delta x$, as indicated in 
the legend (from Ref.~\cite{PX}).}
\end{center}
\end{figure}

This behavior of the steady-state spatial persistence probability
has been confirmed~\cite{PX} from simulations of the Family model.
Two length scales have to be taken into consideration in the
interpretation of the numerical results:
the size $L$ of the sample used in the
simulation, and the sampling distance $\delta x$ which denotes the spacing
between two successive points where the height variables are measured in the
calculation of the persistence probability. The minimum value of $\delta x$
is obviously one lattice spacing, but one can use a larger integral value
of $\delta x$ in the calculation of persistence and survival probabilities.
Examples of the effects of finite $L$ and $\delta x$ 
in simulation results for the Family model are shown in Fig.~\ref{fig10}.
The plots in panel (a) show that for $P_{SS}(x)$ measured in
systems with different sizes, using the smallest
possible value for $\delta x$ (i.e. $\delta x=1$), the exponent
associated with the power-law decay of the persistence probability does
not change, but there is an abrupt downward 
departure from power-law behavior
near $x=L/2$. In panel (b),
we have shown the results for $P_{SS}(x)$
when $L$ remains fixed and $\delta x$ is varied.
Since the persistence
probability is, by definition, equal to unity for $x=\delta x$,
we have plotted $P_{SS}$
as a function of $x/\delta x$ in this figure to ensure that the plots for
different values of $\delta x$ coincide for small values of the $x$-coordinate.
The plots for different $\delta x$ are found to splay away from each other
at large values of $x/\delta x$, with the plots for larger $\delta x$
exhibiting more pronounced downward bending. The numerical results~\cite{PX}
indicate that the dependence of $P_{SS}(x)$ on $L$ and $\delta x$ exhibits a 
scaling behavior similar to that found for the temporal persistence and
survival probabilities:
\begin{equation}
P_{SS}(x,L,\delta x) = f_{SS}(x/L,~ \delta x/L),
\label{f1}
\end{equation}
\noindent where the function $f_{SS}(x_{1},x_{2})$ shows power-law decay
with exponent $\theta_{SS}$ as a function of $x_1$ for small values of
$x_1$ and $x_2 \ll 1$. Fits of the numerical data to a power-law yields
$\theta_{SS}\simeq 0.51$, in good agreement with the expected
value of $1/2$. The probability of persistent large deviations of spatial
fluctuations and the associated family of exponents can
be defined in analogy with their temporal counterparts. These exponents
for 1d EW interfaces have been obtained in Ref.\cite{map} 
using simulations of the Family model. As expected, the dependence of the
spatial persistent large deviations exponent on the parameter $s$ is
found to be identical to that of the temporal persistent large 
deviations exponent for 1d Brownian motion.

Numerical results obtained in Ref.\cite{PX} indicate that 
for 1d EW interfaces at equilibrium, the dependence of the
steady-state spatial survival probability $S_{SS}(x)$ on $x$ is not
described by exponential or power-law forms over a large range of
$x$ values. Simulations for different sample sizes 
and different values of the sampling interval $\delta x$ reveal
that the survival probability is a function of the scaling variable $x/L$,
and that its dependence on $\delta x$ is weak. These numerical results have 
been explained in a recent analytic study~\cite{ssp} in which one of the
authors was involved. This study makes use of the exact mapping of the
spatial statistics of 1d EW interfaces at equilibrium to
the temporal statistics of 1d Brownian motion. The effects
of periodic boundary conditions used in the simulations and the fact that
$h(x,t)$ is measured relative to its instantaneous spatial average (so
that the integral of $h(x,t)$ over $x$ from $x=0$ to $x=L$ 
is strictly equal to zero) are
taken into account in this calculation. Using an exact path integral
formulation, it has been shown 
that $S_{SS}(x,L)$ is a function of $x/L$, and an exact expression for this
function in term of complicated integrals
that can, in principle, be evaluated has been obtained. 
A simpler closed-form expression for this 
function was also obtained in Ref.~\cite{ssp} 
from a simple ``deterministic'' approximation
and it was shown that the results 
obtained from this calculation agree very well with
those of simulations.  

\begin{figure}
\begin{center}
\includegraphics[height=10cm,width=7cm,angle=270]{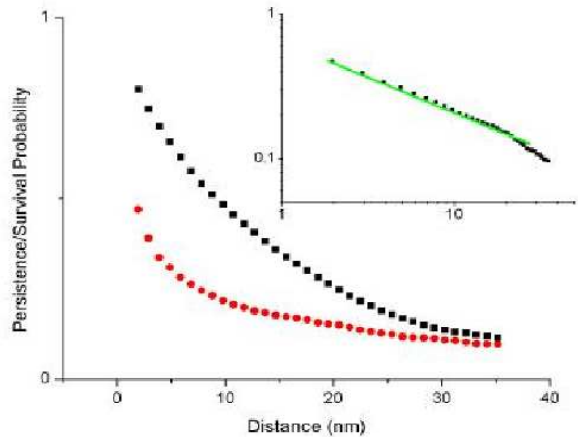}
\caption{\label{fig11} 
Representative spatial persistence and survival probability data.
The data were taken at 970 K, from an STM image with pixel size of
0.977 nm. The persistence and survival probabilities are represented by squares
and circles respectively. The inset is the same persistence curve using
logarithmic scales. The solid green line is a power law fit to the data
over the linear region with the
steady-state spatial persistence exponent $\theta_{SS} = 0.59$ (from
Ref.~\cite{spexp}).}
\end{center}
\end{figure}

Spatial first-passage properties of equilibrium 
step fluctuations governed by the
AD mechanism have also been measured~\cite{spexp} for the Al/Si(111) system.
As mentioned earlier, experimental systems display a
dramatic temperature dependence due to thermally activated kinetics.
As a result, it is possible to decrease the rate of step fluctuations
to an immeasurably slow value, yielding a static spatial step structure
that represents an equilibrium configuration frozen in time. 
For the Al/Si(111) system, fluctuations are essentially absent over time
intervals of several minutes for temperatures below
770 K~\cite{lyubin2}, while at 1020 K steps fluctuate with times scales
of the order of seconds~\cite{lyubin}. Therefore, to obtain viable information
above 870 K, samples were prepared at elevated temperatures and
were then quenched at an initial cooling rate of over 200 K/s to
room temperature in order to capture and preserve the step-edge
displacements~\cite{lyubin2}.
In measurements of spatial first-passage properties,
the spatial separation between adjacent measured points (or pixel size)
plays the role of the sampling distance $\delta x$ mentioned above. 

\begin{figure}
\begin{center}
\includegraphics[height=8cm,width=12cm]{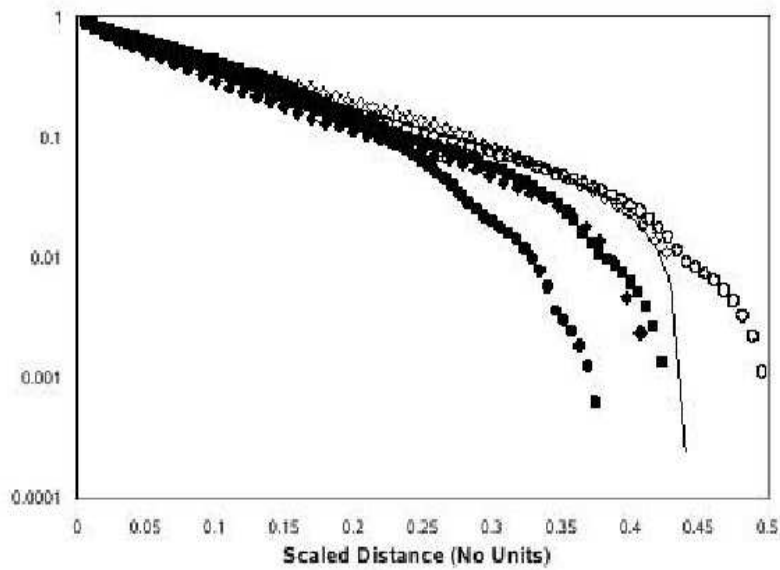}
\caption{\label{fig12} 
Survival probabilities determined from single steps
chosen to display measurements at two different pixel sizes and a wide range
of step lengths.  Solid diamonds, squares, and circles are from 
(500 nm)$^2$ images
and have system lengths of L = 98.9 nm, 170 nm, and 162 nm respectively.
Open diamonds, squares, and circles are from (300 nm)$^2$
images and have system
lengths of L = 65.8 nm, 154 nm, and 87 nm respectively. The survival 
probability is plotted as a function of the scaled distance $x/L$.
The solid line represents the theoretical prediction of 
Ref.\cite{ssp} (from Ref.~\cite{spexp}).}
\end{center}
\end{figure}

Spatial persistence and survival probabilities were measured
for Al/Si(111) surfaces representing spatial equilibrium over
temperatures of 720--1070 K. 
The images used in these measurements were of two sizes,
(300 nm)$^2$ and (500 nm)$^2$, measured with pixel sizes 0.586 nm and 0.977 nm
respectively.  Each step image used for this analysis was cropped from
a larger original STM image, yielding a distribution of effective
system sizes $L$ but the same value of the pixel size $\delta x$.
Experimental data were analyzed to determine both the
spatial persistence and survival probabilities versus distance $x$,
as shown in Fig.~\ref{fig11}. The same persistence
curve is shown in the inset using logarithmic axes to illustrate more clearly
the power-law behavior.  Deviations from the power-law fit
at large distances stem from limited statistics at large $x$, as well
as possible effects of finite measurement size issues.
The average of the persistence curves for all the steps in each
image was fit to a power law to extract the persistence exponent $\theta_{SS}$.
No systematic dependence on temperature was observed, similar to
the lack of temperature-dependence observed for the temporal
persistence.  An analysis of the averaged persistence probabilities
over all the temperatures results in a persistence exponent of
$\theta_{SS} = 0.498 \pm 0.062$, in excellent agreement with the theoretical
value of 1/2.

The measured survival curves $S_{SS}(x)$ showed a great deal of
variability, but scaling the distance $x$ by the length $L$ of the step 
caused the curves to collapse onto one another, 
as illustrated in Fig.~\ref{fig12}.
No systematic effect of the ratio $\delta x/L$ on the linear region 
of the semi-logarithmic plots in Fig.~\ref{fig12} was observed. 
Fits of the scaled survival probabilities to an exponential form gave
good results for short distances ($x/L < 0.2$), 
with an average value of the scaled
decay length $x_s/L = 0.076 \pm 0.033$ with a temperature-dependence
smaller in magnitude than the experimental uncertainty in the data.
The fit of the short-distance survival data to an exponential form
with a fixed decay length is, in fact, consistent with the short-distance
form of the theoretical prediction~\cite{ssp} for the
spatial survival probability of 1d EW interfaces. 
The theoretically predicted
survival probability, shown as the solid line plotted in Fig.~\ref{fig12},
reproduces the rapid fall-off of the survival probability at larger
distances.  Consistent with the experimental observation, the functional
form is indistinguishable from an exponential for $x/L < 0.2$, except
very close to $x=0$ where the theoretical result predicts
a cusp of the form $S_{SS}(u) \sim 1- 4\sqrt{3u}/\pi$ with $u=x/L$.
The empirical survival length
constant $x_s/L$ extracted from exponential fits 
is a useful experimental rule of thumb that
provides a measure of the characteristic fluctuation
length scales relative to the system size.

\section{Summary and discussions} \label{summ} 

The theoretical and experimental results summarized above provide a fairly 
complete description of temporal and spatial first-passage properties of 
the equilibrium fluctuations of isolated steps on a surface. The close
agreement between theoretical (analytic when available and numerical) 
and experimental results in nearly all the
cases studied provides a confirmation of 
several theoretical ideas about first-passage
properties of dynamical fluctuations of spatially extended objects, validates
the simple theoretical models used to describe these fluctuations, and 
establishes the usefulness of first-passage statistics in characterizing the
nature of these fluctuations. Due to the non-Markovian nature of these
fluctuations, it is   
difficult to carry out analytic calculations of some of 
the first-passage probabilities
considered here. In particular, no analytic theory is available at 
present for the family of persistent large deviations exponents $\theta_l(s)$
and the time scales for the decay of the generalized survival probability
$S(t,R)$ and the generalized inside survival probability $S_{in}(t,R)$. 
The development of analytic
theories for these quantities would be very interesting and useful. 

The results reviewed above also indicate that in general, 
the issues of discrete sampling, finite measurement time, and
finite effective step length make the interpretation
of non-universal aspects of
persistence and survival data for step fluctuations a difficult prospect.
These difficulties do not detract from the remarkably
successful interplay between
experiment and theory that has enhanced fundamental
understanding of the statistical
physics of model 1d interfaces.
Nevertheless, since first-passage statistics
speaks to eminently practical concerns related to structural stability,
the interpretation of measurements in a material and
temperature dependent context
is crucial. We have shown that the dependence of first-passage 
probabilities on sampling parameters and effective step length can be
quantified in terms of phenomenological scaling relations obtained using
plausible arguments and confirmed from numerical studies. 
The development of a deeper (analytic if possible) understanding
of these scaling relations would be very useful in extracting quantitative
information from experimental studies of first-passage properties.

The studies described here may be extended in several directions. 
Most nonlinear 
models~\cite{KPZ,DT,WV,chandan} used to describe the  nonequilibrium kinetics
of surface growth do not have the $h \to -h$ symmetry of the linear models
considered here. This lack of symmetry translates into
a difference between the exponents~\cite{PT,Krug2} that describe the decay of 
steady-state temporal persistence
probabilities for positive and negative displacements of the 
interface from its initial position. The smaller of these two 
persistence exponents is known~\cite{PT} to be equal to $(1-\beta)$ where 
$\beta$ is the dynamical growth exponent of the nonlinear 
model, but no analytic result is available for the other exponent. Also, it
has been found in simulations~\cite{PT,Krug1,Krug2} 
that ``transient'' persistence exponents that describe the temporal 
first-passage statistics in
the initial regime of interface growth from a flat initial state are different
from the steady-state persistence exponents. Studies of 
some of the other first-passage properties considered in our work for these
situations would be interesting. Other interesting questions that deserve
attention include the effects of step-step interactions on the first-passage
properties of step fluctuations and persistence and survival properties of
higher-dimensional interfaces such as fluctuating membranes and growing
surfaces in two dimensions.

\ack

We would like to thank 
O. Bondarchuk, B. R. Conrad, W. G. Cullen, M. Degawa, I. L. Lyubinetsky, 
S. N. Majumdar, P. Punyindu 
Chatraphorn and C. G. Tao for their contributions in some of the 
studies described here. 
We would also like to thank S. N.
Majumdar and J. Krug
for helpful discussions. CD would like to thank
the Condensed Matter Theory Center, Laboratory for Physical Sciences 
of the University of Maryland for support
and hospitality during a sabbatical visit. This work was supported by 
the Condensed Matter Theory Center of the University of Maryland and 
UMD NSF-MRSEC under Grant No. DMR 05-20471.


\section*{References}

\begin{thebibliography}{999}
%
%
\bibitem{redner_book} S. Redner, {\it A Guide to First-passage
Processes} (Cambridge University Press, Cambridge, 2001).
%
\bibitem{pers_review} S. N. Majumdar, Curr. Sci. {\bf 77}, 370 (1999).
%
\bibitem{diffusion} S. N. Majumdar, C. Sire, A. J. Bray, and
S. J. Cornell, Phys. Rev. Lett. {\bf 77}, 2867 (1996);
B. Derrida, V. Hakim, and R. Zeitak, Phys. Rev. Lett.
{\bf 77}, 2871 (1996).
%
%
%
\bibitem{spin1} B. Derrida, V. Hakim, and V. Pasquier, Phys. Rev. Lett.
{\bf 75}, 751 (1995).
%
\bibitem{spin2} S. N. Majumdar and C. Sire, Phys. Rev. Lett.
{\bf 77}, 1420 (1996).
%
\bibitem{critical} S. N. Majumdar, A. J. Bray, S. J. Cornell and C. Sire,
Phys. Rev. Lett. {\bf 77}, 3704 (1996).
%
\bibitem{fisher} D.S. Fisher, P Le Doussal and C. Monthus, Phys. Rev. Lett.
{\bf 80}, 3539 (1998).
%
\bibitem{stock} M. Constantin and S. Das Sarma, Phys. Rev. E {\bf 72}, 051106
(2005).
%
%
\bibitem{droplet} M. Marcos-Martin, D. Beysens, J. P. Bouchaud,
C. Godreche, and I. Yekutieli, Physica A {\bf 214}, 396 (1995).
%
\bibitem{soap} W. Y. Tam, R. Zeitak, K. Y. Szeto, and J. Stavans,
Phys. Rev. Lett. {\bf 78}, 1588 (1997);
W. Y. Tam and K. Y. Szeto, Phys. Rev. E. {\bf 65}, 042601 (2002).
%
\bibitem{liqcryst} B. Yurke, A. N. Pargellis, S. N. Majumdar, and
C. Sire, Phys. Rev. E {\bf 56}, R40 (1997).
%
\bibitem{Xe} G. P. Wong, R. W. Mair, and R. L. Walsworth,
Phys. Rev. Lett. {\bf 86}, 4156 (2001).
%
\bibitem{px_expt} J. Merikoski, J. Maunuksela, M. Myllys, J. Timonen,
and M. J. Alava , Phys. Rev. Lett. {\bf 90} 024501 (2003).

\bibitem{book1} A. -L. Barabasi and H. E. Stanley, {\it Fractal Concepts
in Surface Growth } (Cambridge University Press, Cambridge, 1995).
%

\bibitem{krugrev} J. Krug, Adv. Phys. {\bf 46}, 139 (1997).
\bibitem{book3} A. Pimpinelli and J. Villain, {\it Physics of Crystal
Growth} (Cambridge University Press, Cambridge, 1998).
%
\bibitem{misbah} C. Misbah, O. Pierre-Louis, and Y. Saito, to be
published (2007).
%
\bibitem{ellen} H.-C. Jeong and E. D. Williams, Surf. Sci. Reports
{\bf 34}, 171 (1999).
%
\bibitem{giesen} M. Giesen, Prog. Surf. Sci. {\bf 68}, 1 (2001).
\bibitem{Dan} D. B. Dougherty, I. L. Lyubinetsky, E. D. Williams,
M. Constantin, C. Dasgupta, and S. Das Sarma,
Phys. Rev. Lett {\bf 89}, 136102 (2002).
%
\bibitem{alex} D. B. Dougherty, O. Bondarchuk, M. Degawa, and
E. D. Williams, Surf. Sci. {\bf 527} L213 (2003).
%
\bibitem{PTS} M. Constantin, S. Das Sarma, C. Dasgupta, O. Bondarchuk,
D. B. Dougherty, and E. D. Williams, Phys. Rev. Lett.
{\bf 91}, 086103 (2003).
%
\bibitem{ST} C. Dasgupta, M. Constantin, S. Das Sarma, and S. N. Majumdar,
Phys. Rev. E {\bf 69}, 022101 (2004).
%
\bibitem{STS} M. Constantin and S. Das Sarma, Phys. Rev. E
{\bf 69}, 052601 (2004).
%
\bibitem{PT} M. Constantin, C. Dasgupta, P. Punyindu
Chatraphorn, S. N. Majumdar, and S. Das Sarma, Phys. Rev. E
{\bf 69}, 061608 (2004).

\bibitem{PX} M. Constantin, S. Das Sarma, and C. Dasgupta,
Phys. Rev. E {\bf 69}, 051603 (2004).

\bibitem{map} M. Constantin and S. Das Sarma,
Phys. Rev. E {\bf 70}, 041602 (2004).

\bibitem{Dan2} D. B. Dougherty, C. Tao, O. Bondarchuk, W. G. Cullen,
E. D. Williams, M. Constantin, C. Dasgupta, and S. Das Sarma,
Phys. Rev. E {\bf 71}, 021602 (2005).
%
\bibitem{expt3}
O. Bondarchuk, D. B. Dougherty, M. Degawa, E. D. Williams,
M. Constantin, C. Dasgupta, and S. Das Sarma,
Phys. Rev. B {\bf 71}, 045426 (2005).

\bibitem{ssp} S. N. Majumdar and C. Dasgupta, Phys. Rev. E {\bf 73},
011602 (2006).

\bibitem{spexp} B. R. Conrad, W. G. Cullen, D. B. Dougherty, 
I. Lyubinetsky, E. D. Williams, Physical Rev. E, in press (2007). 

\bibitem{unpub} C. G. Tao, W. G. Cullen, E. D. Williams and C. Dasgupta,
submitted (2007).
%
\bibitem{Krug1} J. Krug, S. N. Majumdar, S. J. Cornell, A. J. Bray, and
C. Sire, Phys. Rev. E {\bf 56}, 2702 (1997).
%
\bibitem{Krug2} H. Kallabis and J. Krug, Europhys. Lett
{\bf 45}, 20 (1999).
%
\bibitem{Krug4} J. Krug, Physica A {\bf 340}, 647 (2004).
%
\bibitem{nature} V.V. Zhirnov and R.K. Cavin, Nature Materials {\bf 5},
11 (2006).
%
\bibitem{science} G. K. Ramachandran, T. J. Hopson, A. M. Rawlett, L. A.
Nagahara, A. Primak, and S. M. Lindsay, Science {\bf 300}, 1413 (2003).
%
\bibitem{science2} B. Q. Xu and N. J. J. Tao, Science {\bf 301}, 1221 (2003).
%
\bibitem{Dornic1} I. Dornic and C. Godreche, J. Phys. A
{\bf 31}, 5413 (1998).
%
\bibitem{zoltan} T. J. Newman and Z. Toroczkai, Phys. Rev. E
{\bf 58}, R2685 (1998);
Z. Toroczkai, T. J. Newman, and S. Das Sarma, Phys. Rev. E
{\bf 60}, R1115 (1998).

\bibitem{bartelt} N. C. Bartelt, J. L. Goldberg, T. L. Einstein,
and Ellen D. Williams, Surf. Sci. {\bf 273}, 252 (1992).

\bibitem{EW} S. F. Edwards and D. R. Wilkinson, Proc. R. Socs. London,
Ser.A {\bf 381}, 17 (1982).
%
\bibitem{MH} W. W. Mullins, J. Appl. Phys. {\bf 28}, 333 (1957);
C. Herring, J. Appl. Phys. {\bf 21}, 301 (1950).
%
\bibitem{th1} S.V.  Khare and T.L.Einstein,  Phys. Rev. B {\bf 57}, 
4782 (1998).
%
\bibitem{th2} T. Ihle, C. Misbah, and O. Pierre-Louis, Phys. Rev. B {\bf 58}, 
2289 (1998).
%
\bibitem{F} F. Family, J. Phys. A: Math. Gen. {\bf 19}, L441 (1986).
%
\bibitem{racz} Z. Racz, M. Siegert, D. Liu, and M. Pliscke,
Phys. Rev. A {\bf 43}, 5275 (1991).

\bibitem{lyubin} I. L. Lyubinetsky, D. B. Dougherty, T. L. Einstein,
and E. D. Williams Phys. Rev. B {\bf 66}, 085327 (2002).

\bibitem{slepian} D. Slepian, Bell Syst. Tech. J.
{\bf 41}, 463 (1962).

\bibitem{FBM} B. B. Mandelbrot and J. W. van Ness, SIAM Rev. 
{\bf 10}, 422 (1968).

\bibitem{dan3}  D.B. Dougherty, I. Lyubinetsky, T.L. Einstein, 
and E.D. Williams, Phys. Rev. B {\bf 70}, 235422 (2004).

\bibitem{dan6} L. Kuipers, M.S. Hoogeman, J.W.M. Frenken, and H. van Beijeren,
Phys. Rev. B {\bf 52}, 11387 (1995). 

\bibitem{dan7} S. Speller, W. Heiland, A. Biederman, E. Platzgummer, 
C. Nagl, M. Schmid, and P.
Varga, Surf. Sci. {\bf 331}, 1056 (1995). 

\bibitem{th_Pts} A. Baldassarri, J. P. Bouchaud, I. Dornic, and
C. Godreche, Phys. Rev. E {\bf 59}, R20 (1999).

\bibitem{pers} G. F. Newell and M. Roseblatt, Ann. Math. Stat.
{\bf 33}, 1306 (1962).

\bibitem{deltat} S. N. Majumdar, A. J. Bray and G. C. M. A. Ehrhardt, Phys.
Rev. E {\bf 64}, 015101 (2001).

\bibitem{samp} G. C. M. A. Ehrhardt, A. J. Bray, and S.
N. Majumdar, Phys. Rev. E {\bf 65}, 041102 (2002).


\bibitem{maj1} S. N. Majumdar and A. J. Bray, Phys. Rev. Lett. {\bf 86},
3700 (2001).

\bibitem{lyubin2} I. Lyubinetsky, D. B. Dougherty, H. L. Richards,
T. L. Einstein, and E. D. Williams, Surf. Sci. {\bf 492}, L671 (2001).
%
\bibitem{KPZ} M. Kardar, G. Parisi, and Y. -C. Zhang, Phys. Rev. Lett.
{\bf 56}, 889 (1986).

\bibitem{DT} S. Das Sarma and P. Tamborenea, Phys. Rev. Lett.
{\bf 66}, 325 (1991); P. Tamborenea and S. Das Sarma, Phys. Rev. E
{\bf 48}, 2575 (1993).
%
\bibitem{WV} D. Wolf and J. Villain, Europhys. Lett. {\bf 13}, 389 (1990).
%
\bibitem{chandan} C. Dasgupta, S. Das Sarma, and J. M. Kim, Phys. Rev. E
{\bf 54}, R4552 (1996);
C. Dasgupta, J. M. Kim, M. Dutta, and S. Das Sarma, Phys. Rev. E
{\bf 55}, 2235 (1997).
\end{thebibliography}
\end{document}